\documentclass[showpacs,aps,prd,floatfix,amsmath,amssymb]{revtex4}
\usepackage{graphicx}
\usepackage{epstopdf}
\usepackage[utf8]{inputenc}

\begin{document}
\title{Weak dipole moments of the tau lepton in models with an extended scalar sector}
\author{M. A. Arroyo-Ure\~na}
 \affiliation{Facultad de Ciencias F\'{i}sico Matem\'aticas, Benem\'erita Universidad Au\-t\'o\-no\-ma de Puebla,
 Apartado postal 1152, 72001 Puebla, Pue., M\'exico.}

\author{
G. Hern\'andez-Tom\'e}
\affiliation{
Departamento de F\'{\i}sica, CINVESTAV
IPN, Apartado Postal 14-740, 07000, M\'exico D. F., M\'exico.
}

\author{G. Tavares-Velasco}
 \affiliation{Facultad de Ciencias F\'{i}sico Matem\'aticas, Benem\'erita Universidad Au\-t\'o\-no\-ma de Puebla,
 Apartado postal 1152, 72001 Puebla, Pue., M\'exico.}

\begin{abstract}
We consider   renormalizable couplings of  neutral, singly, and doubly charged scalar bosons  to leptons and the $Z$ gauge boson and calculate the one-loop contributions  to the  anomalous weak magnetic dipole moment (AWMDM) $a_\tau^W$   and the   weak electric dipole moment (WEDM) $d_\tau^W$  of a charged lepton in a model independent way. The analytic expressions  are presented in terms of both parametric integrals and Passarino-Veltman scalar functions.  Among the new contributions, there are those arising from the vertices of the type $\phi^\pm W^\mp Z$ and $Z \phi_i\phi_j$ ($i\ne j$), along with contributions from doubly charged scalar bosons. Both $a_\tau$ and $d_\tau^W$ are evaluated in several scenarios , first in a model independent way and then within some popular models, such as  two-Higgs doublet models (THDMs), multiple-Higgs doublet models and Higgs triplet models. As far as  $a_\tau^W$ is concerned, its real part reaches values as high as $10^{-10}-10^{-9}$ for masses of the new scalar bosons in the 200 GeV range, whereas the imaginary part  is one or two orders of magnitude below. On the other hand,  the most promising scenario for a nonvanishing WEDM is offered by a $CP$-violating THDM in a scenario where the heavy neutral scalar bosons  are a mixture of $CP$ eigenstates. It is found that the real part of $d_\tau^W$ is of the order of $10^{-24}$ ecm and its imaginary part can reach the $10^{-26}$ ecm level for masses of the new scalar bosons of the order of a few hundred of GeVs. Both the tau AWMDM and WEDM decrease dramatically as the scalar boson mass increase.

\end{abstract}

 \date{\today}
\preprint{}
\maketitle
\section{Introduction}
\label{Intro}
After the 2012 discovery of the Higgs boson by the ATLAS and CMS collaborations at the  CERN LHC \cite{Aad:2012tfa,Chatrchyan:2012xdj}, the precise determination of this particle's properties has become  one of the most expedite tasks for the experimentalist and so is  the search for new physics (NP) effects, which may help us to shed light on the yet unanswered questions of the standard model  (SM). The mechanism of spontaneous symmetry breaking (SSB) is achieved in the SM by one complex $SU(2)_L$ scalar doublet, thereby leaving as remnant only one physical Higgs boson. However, there is no compelling reason  to expect that this minimal Higgs sector  is the one realized in nature. The most simple  SM extensions are obtained when  one or more scalar multiplets are added to the usual SM Higgs doublet, thereby increasing the spectrum  of physical scalar bosons. Therefore,  models with an extended scalar sector stand out among the most popular and simple SM extensions. A key issue to construct this class of models is to satisfy the $\rho\simeq 1$ relation, along with other theoretical and experimental constraints.  It is well known that Higgs multiplet models containing $N$ multiplets with isospin $T_i$ and hypercharge $Y_i$, whose neutral components develop vacuum expectation values (VEVs) $v_i$, modify the tree-level $\rho$ parameter value  as follows \cite{Gunion:1989we}

\begin{equation}
\label{rhoparameter}
\rho=\frac{\sum_i^N c_i(T_i(T_i+1)-\frac{Y_i^2}{4})v_i^2}{\sum_i^N\frac{Y_i^2 v_i^2}{2}},
\end{equation}
where $c_i=1/2$ (1) for real (complex) multiplets. Therefore, only those models with an extended scalar sector  satisfying the $\rho\simeq 1$ relationship without invoking intricate assumptions are phenomenologically interesting.

Particularly interesting among the models with an extended scalar sector that obey the $\rho=1$ relation at the tree level are  Higgs singlet models (HSMs), two-Higgs doublet models (THDMs), multiple-Higgs doublet models (MHDMs), and some specific  Higgs triplet models (HTMs). Apart from their simplicity, this class of models have several motivations:   new sources of $CP$ violation, the presence of a dark matter candidate, the possibility of accomplishing the see-saw mechanism, the appearance of doubly charged scalar bosons, new tree-level scalar-to-gauge boson couplings, etc. In addition, although these models are interesting by their own,  they can be required by other more sophisticated SM extensions, such as  the minimal supersymmetric standard model (MSSM), whose scalar sector is a THDM.

NP  effects can be searched for indirectly through virtual corrections from new particles predicted by  SM extensions. Along this line, the study of the static electromagnetic properties of fermions provides a unique  opportunity to search for  NP effects. The theoretical study of both the anomalous magnetic dipole moment (AMDM) and the electric dipole moment (EDM) of fermions has long received considerable attention, which   has been boosted  in recent years due to  the  significant progress in the experimental area. After the study of the electromagnetic properties  of a fermion, there has also been great interest in  its static weak properties, which  are associated with its interaction with the $Z$ gauge boson. The analogues of the AMDM and the EDM  are  the anomalous weak magnetic dipole moment (AWMDM)   $a_{f}^{W}$ and the weak electric dipole moment (WEDM)  $d_{f}^{W}$, respectively,   which are defined at the $Z$-pole via the dipole terms of the $Z\bar{f}f$   vertex function
\begin{equation}
ie\bar{u}(p)\Gamma_{Z\bar{f}f}^{\mu}\left(q^{2}\right)u(p')=ie\bar{u}(p)\left(
F_{2}\left(q^{2}\right)i\sigma^{\mu\nu}q_{\nu}+F_{3}\left(q^{2}\right)\sigma^{\mu\nu}\gamma_{5}q_{\nu}  \right)u(p')
\end{equation}
where   $q=p-p'$ is  the $Z$ transfer momentum.  The AWMDM is defined as $a_{f}^{W}=-2m_{f}F_{2}\left(m_{Z}^{2}\right)$  and   the WEDM is given by $d_{f}^{W}=-eF_{3}\left(m_{Z}^{2}\right)$. In the SM,   $a_{f}^{W}$ arises at the one-loop level and $d_f^W$ is induced up to the three-loop level \cite{Booth:1993af}.   Only the weak dipole moments (WDMs) of heavy fermions are worth studying as those of lighter fermions would be beyond the reach of experimental detection.
For instance, in the SM  $a_{\tau}^W=-(2.10 + 0.61i)\times 10^{-6}$ \cite{Bernabeu:1994wh} and  $d_{\tau}^W< 8\times 10^{-34}$ ecm  \cite{Bernreuther:1988jr}.  Although the sensitivity reached at the LEP  was beyond such a precision level,   potentially large contributions  from  SM extensions can be at the reach of future experiments.  The current bounds on the static weak properties of the tau  lepton, which were obtained through the study of $\tau^+\tau^-$ production at the LEP by the ALEPH collaboration \cite{Heister:2002ik}, which used a data sample collected from 1990 to 1995 corresponding to an integrated luminosity of $155$ pb$^{-1}$, are shown in Table \ref{WDMbounds}.  These bounds are well beyond the sensitivity required to test the SM predictions and it is thus worth studying the NP contributions as they could be large enough to be at the reach of detection in the future.

\begin{table}[htb!]
\begin{center}
\caption{Experimental upper bounds on the static weak properties of the tau lepton \cite{Heister:2002ik}.\label{WDMbounds}}
\begin{tabular}{llll}
\hline
\hline
&Real part&&Imaginary part\\
\hline
\hline
$a^W_\tau$&$1{.}1\times10^{-3}$&&$2{.}7\times10^{-3}$\\
\hline
$d^W_\tau$(ecm)&$0{.}5\times10^{-17}$&&$1{.}1\times10^{-17}$\\
\hline
\hline
\end{tabular}
\end{center}
\end{table}

The AWMDM and WEDM of a fermion have been studied in the context of THDMs \cite{Bernabeu:1995gs,GomezDumm:1999tz,Arroyo-Urena:2015uoa}, supersymmetric theories \cite{Hollik:1998wk,Hollik:1998vz}, unparticles \cite{Moyotl:2012zz}, leptoquarks \cite{Bolanos:2013tda}, and the simplest little Higgs model \cite{Arroyo-Urena:2016ygo}.  In this work we are interested in analyzing the new  contributions  arising from  models with an extended scalar sector. We will thus calculate  the one-loop contributions   induced  by  neutral, singly and doubly charged scalar bosons.  Our calculation and numerical analysis will be performed in a model-independent fashion, afterwards we will discuss the possible implications of some specific models with an extended scalar sector, such as  SHMs, THDMs, MHDMs, HTMs, and other models with exotic Higgs sectors.    Our results will also be useful to compute the contributions arising from the scalar sector of models with an extended gauge sector, which also require additional Higgs multiplets, such as  the MSSM, little Higgs models, left-right symmetric models (LRSM) \cite{Mohapatra:1974gc},  331 models \cite{Pisano:1991ee,Frampton:1992wt}, etc.

The  rest of this work is organized as follows. A model-independent calculation of the contribution of new scalar particles to the static weak dipole moments of a lepton is presented in Section \ref{Calculation}. In Section \ref{NumAnalysis} we present the numerical analysis of the AWMDM and WEDM of the tau lepton,  along with the implications for the contributions of  models with  extra scalar multiplets. The conclusions and outlook are presented in Section \ref{Conclusions}. Finally, the necessary Feynman rules and some lengthy formulas for the loop integrals are presented in the Appendices.

\section{New scalar contributions to the AWMDM and WEDM of charged leptons}
\label{Calculation}

We are interested in the contributions to the AWMDM and WEDM of charged leptons  from new neutral, charged and doubly charged scalar particles, which can arise in several models with an extended scalar sector.  Our  calculation will thus be somewhat general: instead of working out the weak dipole moments (WDMs) within a specific model, we will  consider  the scenario of a theory with several nondegenerate neutral, singly and doubly charged scalar bosons with the most general renormalizable couplings to the leptons and the $Z$ gauge boson that can induce the WDMs at the one-loop level. Once our model-independent calculation is presented via  the unitary gauge, we will perform the numerical analysis and consider the implications of a few models.

\subsection{Contributions from new neutral and singly charged scalar bosons}
We first consider lepton number conserving (LNC) interactions mediated by scalar bosons [lepton number violating interactions (LNV) can be induced by the doubly charged scalar bosons]. For the couplings of a lepton-antilepton pair with a neutral or singly charged scalar particle (denoted $\phi_i$ or $\phi_j$ from now on) we will consider the following renormalizable interaction

\begin{equation}
\mathcal{L}=ig\,\bar{\ell}_{l}\left(S_{ilm}+P_{ilm}\gamma_{5}\right)\ell_{m}\phi_{i}+\textrm{H.c}. ,\label{Hff}
\end{equation}
where $\ell_{l}$ is a charged lepton and $\ell_{m}$ is a lepton whose charge depends on that  of the scalar boson: if $\phi_{i}$ is a neutral (charged) scalar boson, $\ell_m$ is a charged  (neutral) lepton. Also, note that we will introduced a factor of $g$ for each coupling,  we thus expect  that the  $S_{ilm}$ and $P_{ilm}$ couplings are of the order of $O(1)$ or lower. Note also that we are considering the most general case where the neutral scalar bosons are a mixture of $CP$-even and $CP$-odd states, which can arise for instance in   THDMs with $CP$ violation.

As for the interactions  of $Z$ gauge boson with two nondegenerate neutral or  charged scalar bosons $\phi_i$ and $\phi_j$, it will be written as follows
\begin{equation}
\mathcal{L}= i g m_Z\, g_{Z\phi_{i}\phi_{j} }Z^{\mu}\phi_{i}^{\dagger}\overleftrightarrow{\partial_{\mu}}\phi_{j},
\end{equation}
whereas  the couplings of the type $ZV\phi_i$, with $V$ a neutral  (charged) gauge boson and $\phi_i$ a neutral (charged) scalar boson,  can be written as
\begin{equation}
\mathcal{L}= ig\, g_{\phi_{i} VZ}   Z^{\mu}V_{ \mu}\phi_{i}+\textrm{H.c.},
\end{equation}
where $V$ stands for a SM gauge boson or  another one predicted by a SM extension. Such coupling can be for instance the $\phi_i ZZ$ and the $\phi^\pm  Z W^\mp $ ones. The latter can arise in HTMs at the tree-level, whereas in THDMs it arises up to the one-loop level.
We also need  the  interaction between a lepton-antilepton pair with a neutral or charged gauge boson $V$, which we write as

\begin{equation}
\mathcal{L}= ig\,\bar{\ell}_{l}\gamma_{\mu}\left(g^{V l m}_{V} - g^{V l m}_{A}\gamma_{5}\right)V^{\mu}\ell_{m}+\textrm{H.c}.
\end{equation}
Particular expressions for the coupling constants $S_{ilm}$,  $P_{ilm}$, $ g_{Z\phi_{i}\phi_{i} }$, etc., will be  known once a specific model is considered.  Since we are mainly  interested on the contributions arising from  models with an extended scalar sector only,  we will not consider the  contributions  of  hypothetical gauge bosons or fermions  predicted in  SM extensions with an extended gauge sector.

\begin{figure}[htb!]
\begin{center}
\includegraphics[width=10cm]{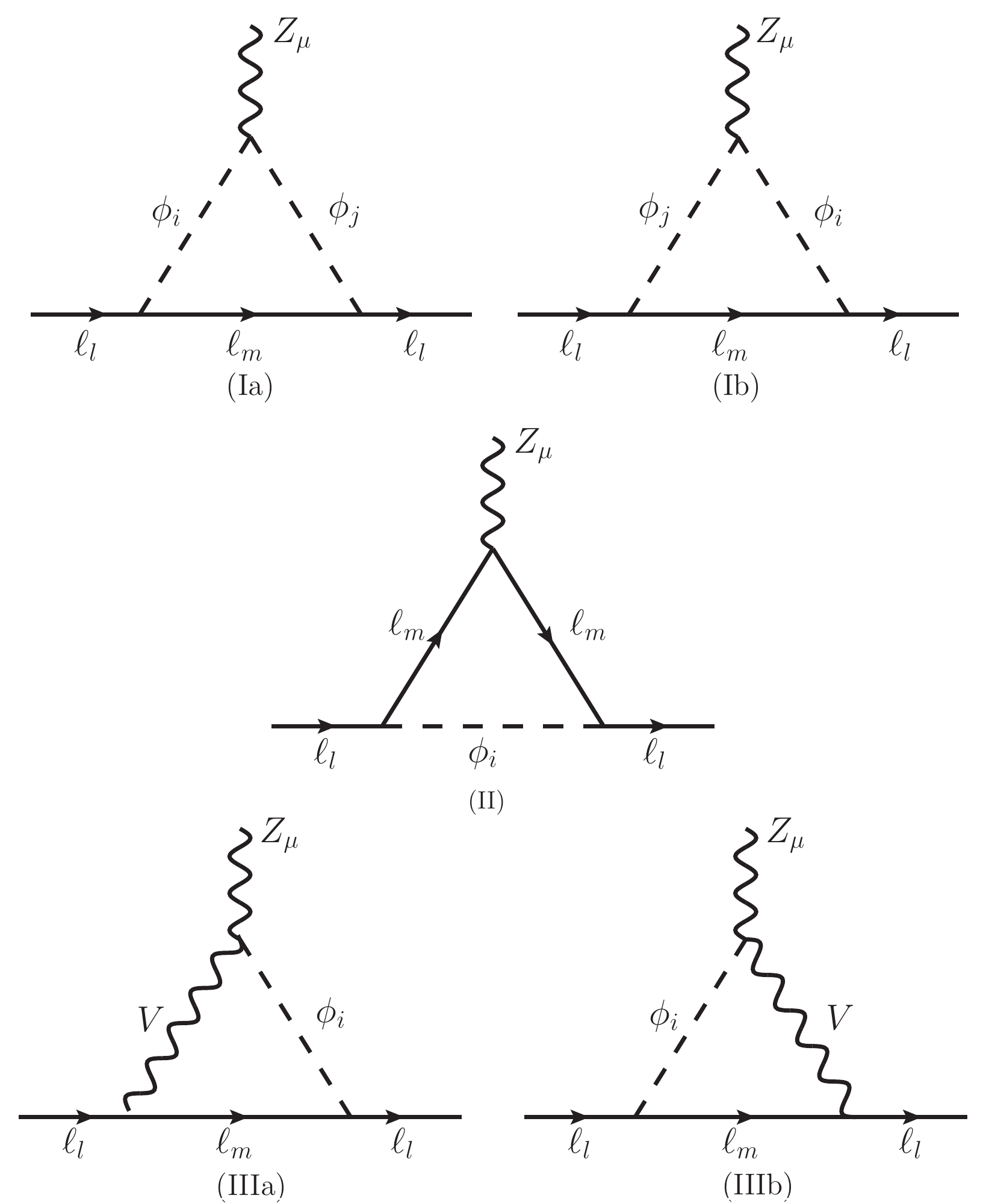}
\caption{Generic Feynman diagrams for the  contributions of new neutral and charged scalar bosons to the AWMDM and WEDM of a charged lepton. Here $\ell_l$ stands for a charged lepton, whereas $\ell_m$ is a lepton whose charge depends on that of the $\phi_{i}$ and $\phi_{j}$   scalar bosons (diagrams I and II) and that of the $V$ gauge boson and the $\phi_i $ scalar boson (diagrams III). } \label{FeynmanDiagrams1}
\par\end{center}
\end{figure}

The Feynman rules for the above described couplings  are presented in Appendix \ref{FeynRules}.  At the one-loop level these couplings
lead to contributions to the WDMs of a charged lepton via the Feynman diagrams depicted in Fig. \ref{FeynmanDiagrams1}, where $\phi_i$ and $\phi_j$ represent neutral or charged scalar bosons, and $V$ is a gauge boson. Evidently once the electric charge of the scalar bosons are fixed, the charges of the internal lepton $\ell_m$ and the gauge boson $V$ will also become fixed by charge conservation in each vertex. For instance,  if $\phi_i$ and $\phi_j$ are  neutral  scalar bosons,   $\ell_m$ is a charged lepton $e, \mu, \tau$. Thus, for the contributions of new neutral scalar bosons we will need  the vertices $\phi_i\ell_m\ell_l$,  $\phi_i ZZ$,  $Z\phi_i\phi_j$ and $Z\bar{\ell}_m\ell_m$. On the other hand,  when $\phi_i$ and $\phi_j$ are charged scalar bosons, the internal lepton is a neutrino $\ell_m=\nu_m$. Therefore, this class of contributions  will require  the vertices $\phi_i^+\bar{\nu}_m\ell_l$,  $\phi_i^-W^+Z$,  $Z\phi_i^-\phi_j^-$ and $Z\bar{\nu}_m\nu_m$.

In order to solve the one-loop integrals we have used both the Feynman parameter technique and the Passarino-Veltman reduction scheme \cite{Passarino:1978jh}, which will allows us to cross check the  results numerically. After some algebra we have obtained  the following  results for the contributions of each type of Feynman diagram to the WDMs of a charged lepton.

\subsubsection{Anomalous weak magnetic dipole moment}

The contributions to the AWMDM can be written as follows

\begin{eqnarray}
a^{W-I}_{ l}&=&\frac{\alpha\sqrt{x_{ l}}}{4\pi s_W^3 }\sum_{i,j, m} 16N_{ij}\ \textrm{Re}\left[S_{ilm}S_{jlm}^{*}g_{Z\phi_{i}\phi_{j}}^{*}\right] A^{ {m}\phi_{i}\phi_{j}}_{I}
   +
  \left(
  \begin{array}{c}
  \sqrt{x_{ m}}\to -\sqrt{x_{ m}}\\
  S_{ilm}\to P_{ilm}\\
  S_{jlm}\to P_{jlm}
  \end{array}
  \right) , \label{MDMI}
\end{eqnarray}

\begin{eqnarray}
a^{W-II}_{ l}&=&\frac{\alpha\sqrt{x_{ l}}}{4\pi s_W^3 }\sum_{i, m}16\left(g_{V}^{Z m  m}\|{S_{i l m}}\|^{2}
 A^{\phi_{i} {m} {m}}_{II_1} +
   g_{A}^{Z m  m}  \sqrt{x_{ l}}\, \textrm{Re}\left[S_{il m}P_{il m}^{*}\right]
  A^{\phi_{i} {m} {m}}_{II_2}\right)  \nonumber\\  &+&
  \left(
  \begin{array}{c}
  \sqrt{x_{ m}}\to -\sqrt{x_{ m}}\\
  S_{il m}\leftrightarrow P_{il m}
  \end{array}
  \right)
  , \label{MDMII}
\end{eqnarray}
and
\begin{eqnarray}
a^{W-III}_{ l}&=&\frac{\alpha\sqrt{x_{ l}}}{4\pi s_W^3 }\sum_{i, m,V}\frac{2 g_{\phi_{i}VZ}}{x_V}
\textrm{Re}\left[S_{ilm}\: g_{V}^{V l  m*}\right] A^{ {m}\phi_{i}V}_{III}
- \left(
  \begin{array}{c}
  \sqrt{x_{ m}}\to -\sqrt{x_{ m}}\\
  S_{ilm}\to  P_{ilm}\\
 g_{V}^{V l  m}\to g_{A}^{V l  m}
  \end{array}
  \right), \label{MDMIII}
\end{eqnarray}
where $x_a=m_a^2/m_Z^2$, $N_{ij}=(1-\delta_{ij})\;(1)$ for neutral (charged) scalar bosons. It is understood that these sums run over all the possible combinations  of internal particles predicted by a particular theory.  The last term is obtained from the first term  after the corresponding replacements are done. As far as the  $A_i^{ABC}$  functions are concerned, they depend on the masses of the particles circulating into each triangular  loop and that of the external lepton (the superscript letters  stand for the distinct particles in the  loop) but such a dependence has not been written out explicitly to avoid  cumbersome equations. The corresponding expressions are presented in Appendix \ref{OneLoopFunctions} in terms of both parametric integrals and Passarino-Veltman scalar functions. At this point, it is worth mentioning some  important aspects of our calculation. Firstly, we have verified that the contributions to both the AWMDM and WEDM from the diagrams of Fig. \ref{FeynmanDiagrams1} are free of ultraviolet divergences. In addition, we have verified that the expressions  (\ref{MDMI}) and (\ref{MDMII}) reduce to those reported in  \cite{Moore:1984eg} for the AMDM of a lepton in the limit of $m_{Z} \to 0$ and after replacing the $Z$ couplings with the photon ones. 

\subsubsection{Weak electric dipole moment}
As for the contributions to the WEDM, they are given as

\begin{eqnarray}
d^{W-I}_{ l}&=&\frac{e \alpha}{4\pi s_W^3  m_Z}\sum_{i, j, m} 4N_{ij}\textrm{Im}\left[P_{ilm}^{*}S_{jlm}g_{Z\phi_{i}\phi_{j}}\right]
D^{ {m}\phi_{i}\phi_{j}}_{I}  +
  \left(
  \begin{array}{c}
  \sqrt{x_{ m}}\to -\sqrt{x_{ m}}\\
    S_{jlm}\to P_{jlm}\\
  P_{ilm}\to S_{ilm}
  \end{array}
  \right) , \label{EDMI}
\end{eqnarray}

\begin{eqnarray}
d^{W-II}_{ l}&=&\frac{e \alpha}{4\pi s_W^3  m_Z}\sum_{i, m}32g_V^{Z  m m}  \, \textrm{Im}\left[S_{ilm}P_{ilm}^{*}\right]
D^{\phi_{i} {m} {m}}_{II},
 \label{EDMII}
\end{eqnarray}
and
\begin{eqnarray}
d^{W-III}_{ l}&=&\frac{e \alpha}{4\pi s_W^3  m_Z}\sum_{i ,m,V}\frac{g_{\phi_{i}VZ}}{x_V}\textrm{Im}
\left[S_{ilm}\: g_A^{V  l m*}\right] D^{ {m}\phi_{i}V}_{III} - \left(
  \begin{array}{c}
  \sqrt{x_{ m}}\to -\sqrt{x_{ m}}\\
  S_{ilm}\to  P_{ilm}\\
 g_A^{V  l m}\to g_V^{V l m}
  \end{array}
  \right),  \label{EDMIII}
\end{eqnarray}
where again the $D^{ABC}_i$ functions are also presented in Appendix \ref{OneLoopFunctions} in terms of both parametric integrals and Passarino-Veltman scalar functions.

\subsection{Contribution from doubly charged scalar bosons}
\begin{figure}[htb!]
\begin{center}
\includegraphics[width=10cm]{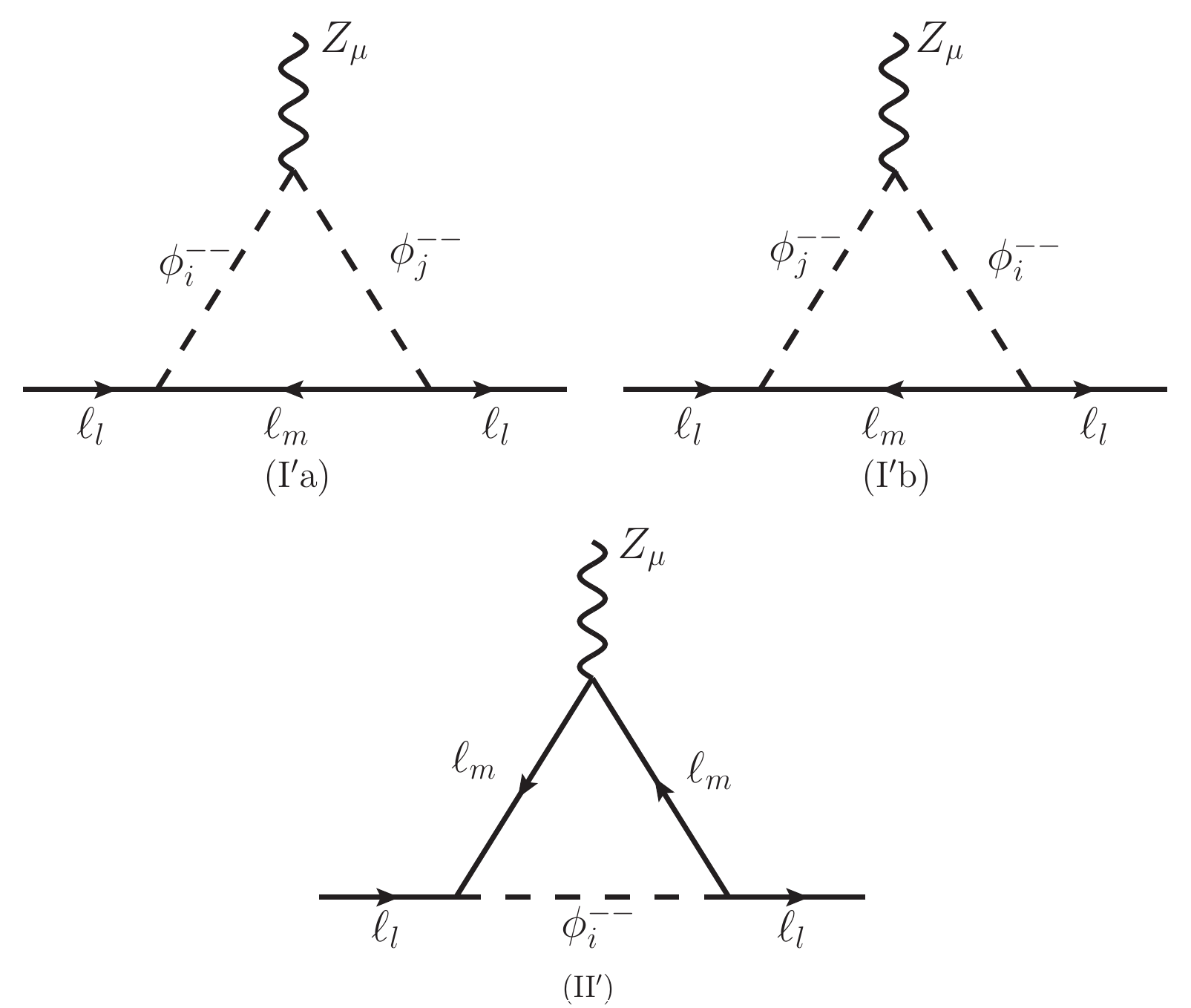}
\caption{Generic Feynman diagrams  for the contributions of doubly charged scalar bosons to the AWMDM and WEDM of charged leptons. Here $\ell_l$ and $\ell_m$ are both charged leptons. } \label{FeynmanDiagram2}
\par\end{center}
\end{figure}

In addition to the above results, we also need to consider the  $\Delta L=2$ lepton number violating (LNV) contributions, which can be mediated by a doubly charged scalar boson. This class of interactions can be written as

\begin{equation}
\mathcal{L}^{\Delta L=2}=
g \,\ell_{l}^{T}C\left(S'_{ilm}+P'_{ilm}\gamma^5\right)\ell_{m}\phi^{i}+\textrm{H.c}\label{LNV}
\end{equation}
where $C$ is the charge conjugation matrix. Doubly charged scalar bosons  can contribute to the AWMDM and WEDM of charged leptons via the Feynman diagrams shown in Fig. \ref{FeynmanDiagram2}, where the fermion-flow arrows either clash or emerge from  LNV vertices as opposed to lepton number conserving vertices, where the fermion flow follows the same direction.  Since we are considering models with an extended scalar sector only, there is no contribution similar to the type-III one, which will require a doubly charged gauge boson. Some models with extended gauge sector, for instance $SU(3)_L\times U(1)_X$ models, predict such particles.

Due to the presence of the charge conjugation matrix and  transposed spinors, the application of the Feynman rules for these diagrams must be done carefully.  We have followed the approach presented in Ref. \cite{Moore:1984eg,Denner:1992vza} for calculating amplitudes with fermion number violating vertices. The   relevant details are presented in Appendix \ref{FeynRules}. After some algebra we have found that the results arising from the Feynman diagrams of Fig. \ref{FeynmanDiagrams1} [i.e. (\ref{MDMI})- (\ref{EDMIII})] hold true for the contributions of a doubly charged scalar bosons except that a factor of two for each LNV vertex must be inserted when the  leptons are identical  ($l=m$):

\begin{eqnarray}
a^{W-I'}_{ l}&= &(1+\delta_{lm})^2 a^{W-I}_{ l},\\
a^{W-II'}_{ l}&= &(1+\delta_{lm})^2 a^{W-II}_{ l},
\end{eqnarray}
where it is understood  that one must replace the appropriate couplings and masses involved in each contribution. Similar expressions hold for the contributions to the WEDM arising from LNV vertices. This situation was also noted in the calculation of the AMDM of a lepton \cite{Moore:1984eg}, which means that the contribution from a doubly charged scalar boson to the AWMDM or  WEDM of a charged lepton is enhanced by a factor of 4 with respect to the singly charged scalar contributions. It is worth noting  that  (\ref{MDMIII}) and (\ref{EDMII}) would be also valid for a type-III' contribution arising from a doubly charged scalar and a doubly charged gauge boson, provided that the factor of 2 for vertices with identical leptons is considered and the respective couplings are used.

We will analyze below  in a model independent way the behavior of the contributions to the AWMDM and WEDM of the tau lepton arising from each type of   contribution.  We will next discuss a few illustrative  scenarios arising in a few models.

\section{Numerical Analysis}
\label{NumAnalysis}

\subsection{Model independent analysis of the AWMDM of the tau lepton}

The possible scenarios  with new  neutral, singly charged, and doubly charged scalar bosons that can induce an AWMDM are summarized in Table  \ref{Couplings}, where we show the corresponding type of Feynman diagram from which such contributions arise and a few specific models in which such a scenario is attained. We will consider that the   neutral scalar bosons are either pure or a mixture of $CP$ eigenstates: $\phi_i^0$ and $\tilde{\phi}_i^0$ will denote  neutral $CP$-even and $CP$-odd scalar bosons, respectively, whereas  $\hat\phi_i^0$ will stand for a mixture of $CP$-eigenstates. The latter can arise for instance in MHDMs with  explicit or spontaneous $CP$ violation in the scalar sector.

As far as neutral scalar bosons are concerned, we note that while  a $CP$-even scalar boson can contribute to the AWMDM through type-II and type-III Feynman diagrams, a $CP$-odd scalar boson can only contribute via the type-II diagram. On the other hand, a pair of nondegenerate $\hat{\phi}_i^0$ and $\hat{\phi}_j^0$  scalar bosons can induce the AWMDM via the three contributions. As for singly charged scalar bosons, they can contribute via both type-I and type-II  contributions, whereas type-III contribution is present only in HTMs, where the $\phi^{\pm}W^\mp Z$ vertex is induced at the tree-level.   Finally, doubly charged scalar bosons can induce the AWMDM via the type-I$'$ and type-II$'$ diagrams, though type-III-like contributions can also be present in models with a doubly charged gauge boson.

\begin{table}
\begin{center}
\caption{Possible contributions to the AWMDM of a charged lepton induced at the one-loop level by new scalar bosons. Here $\phi_{i,j}^0$ ($\tilde{\phi}_{i,j}^0$) stand for neutral $CP$-even ($CP$-odd) scalar bosons and $\hat{\phi}_{i,j}^0$ for a  mixture of $CP$-eigenstates. Note that  although the vertex $Z\phi_i^0\tilde\phi_j^0$ is not forbidden by $CP$ invariance, the type-I contribution to the AWMDM vanishes. \label{Couplings}}
\begin{tabular}{cccc}
\hline
\hline
Scalar boson(s)&Involved couplings&Contribution&Model\\
\hline
\hline
$\phi^0_{i}$&$\phi^0_{i}\bar{\ell}_l\ell_m$, $\phi_{i}^0 ZZ$&II-III&MHDM, TM\\
$\tilde{\phi}^0_{i}$&$\tilde{\phi}^0_{i}\bar{\ell}_l\ell_m$&II&MHDM, TM\\
$\hat\phi^0_{i}$ and $\hat\phi^0_{j}$&$\hat\phi^0_{i, j}\bar{\ell}_l\ell_m$,  $Z\hat\phi_i^0\hat\phi_j^0$ ($i\ne j$),  $\hat\phi_{i, j}^0 ZZ$&I-III&MHDM\\
$\phi^\pm_{i}$&$\phi^-_{i}\bar{\ell}_l\nu_m$,  $Z\phi^{\pm}_i \phi^{\mp}_i$ &I-II&MHDM, TM\\
$\phi^\pm_{i}$&$\phi^-_{i}\bar{\ell}_l\nu_m$,     $ZW^\pm\phi^\mp_{i,j}$&III&TM\\
$\phi^\pm_{i}$ and $\phi^\pm_{j}$&$\phi^-_{i,j}\bar{\ell}_l\nu_m$,  $Z\phi^{\pm}_i \phi^{\mp}_j$ &I-II&MHDM, TM\\
$\phi^{\pm\pm}_{i}$&$\phi^{--}_{i}\ell_l\ell_l$, $Z\phi^{\pm\pm}_i \phi^{\mp\mp}_i$, &I$'$-II$'$& TM\\
$\phi^{\pm\pm}_{i}$ and $\phi^{\pm\pm}_{j}$&$\phi^{--}_{i}\ell_l\ell_l$, $Z\phi^{\pm\pm}_i \phi^{\mp\mp}_j$, &I$'$-II$'$& TM\\
\hline
\hline
\end{tabular}
\end{center}
\end{table}

We now proceed to present a numerical analysis of behavior of the  scalar boson contributions to the AWMDM  of the tau lepton. Our aim is to examine several scenarios for the contributions of neutral, singly charged, and doubly charged scalar bosons. We will first present a model-independent analysis and afterwards we concentrate on some realistic models.

\subsubsection{Neutral scalar bosons}

To assess the potential contributions of new neutral scalar bosons to the AWMDM of the tau lepton we  consider the minimal scenarios for the each type of contribution to be nonvanishing and estimate the corresponding order of magnitude. The minimal scenarios are:
\begin{itemize}
\item Type-I contribution  requires at least two nondegenerate neutral scalars $\hat \phi_1$ and  $\hat \phi_2$ that are a mixture of $CP$ eigenstantes.
\item  Type-II contribution  requires either a single neutral $CP$-even scalar  $\phi^0_1$ or a single neutral $CP$-odd scalar  $\tilde \phi^0_1$.

\item Type-III contribution  can arise via a single  $CP$-even neutral scalar boson $\phi^0_1$.
\end{itemize}

\begin{figure}[htb!]
\begin{center}
\includegraphics[width=13cm]{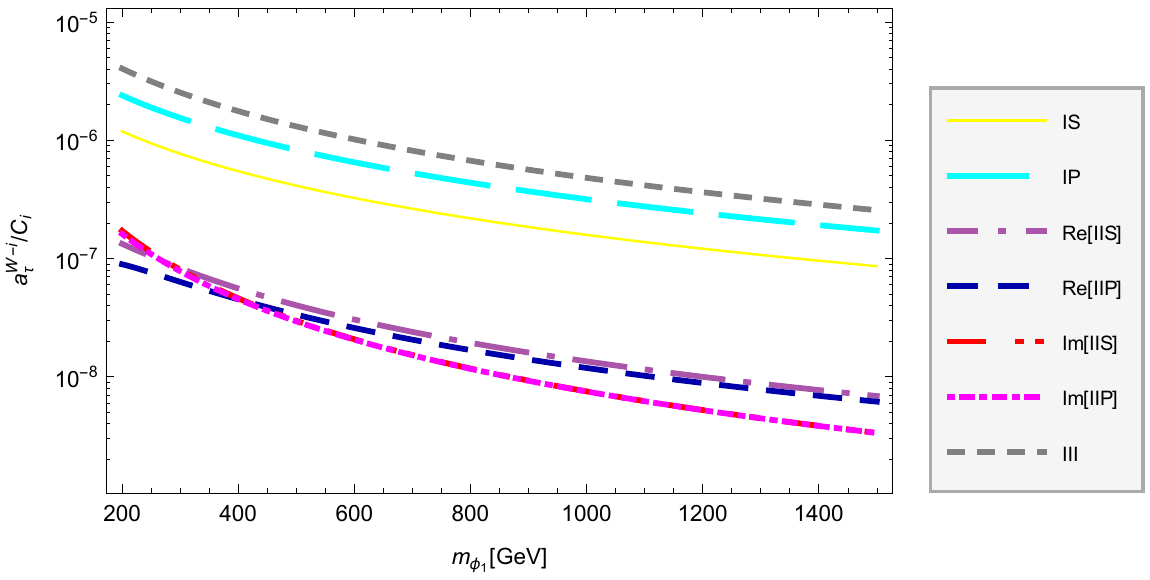}
\caption{Absolute values of the  contributions from new neutral scalar bosons to the AWMDM of the tau lepton  induced by the three types of  Feynman diagrams of Fig. \ref{FeynmanDiagrams1}. Both the real and imaginary parts of type-II contributions are shown. We consider the following scenarios: two nondegenerate scalar bosons  $\hat \phi^0_1$ and $\hat \phi^0_2$ with  $m_{\hat\phi^0_1}=m_{\phi_1}$  and  $m_{\hat\phi^0_2}=200$ GeV (IS and IP), a single $CP$-even  scalar boson $\phi^0_1$ (Re[IIS], Im[IIS], and III), and  a single $CP$-odd scalar boson $\tilde\phi^0_1$ (Re[IIP] and Im[IIP]). For  type-I contributions  we take $S_{i\tau\tau}\gg P_{i\tau\tau}$ (IS) and $P_{i\tau\tau}\gg S_{i\tau\tau}$  (IP), for $i=1,2$. In these scenarios each  kind of contribution is nearly proportional to the following product of coupling constants:
$C_{\rm IS}=g_{Z\phi_1\phi_2} \textrm{Re}\left[S_{1\tau\tau}^{*}S_{2\tau\tau}\right]$,  $C_{\rm IP}=g_{Z\phi_{1}\phi_{2}} \textrm{Re}\left[P_{1\tau\tau}^{*}P_{2\tau\tau}\right]$,  $C_{\rm IIS}=\|{S_{1\tau\tau}}\|^{2}$,   $C_{\rm IIP}=\|{P_{1\tau\tau}}\|^{2}$, and  $C_{\rm III}= g_{\phi_{1}ZZ}
\textrm{Re}\left[S_{1\tau\tau}\; \right]$.  \label{plotneutral}}
\end{center}
\end{figure}

Although there could be lepton flavor violating (LFV) scalar couplings,   they are expected to be more suppressed than lepton flavor conserving (LFC)  couplings and we  neglect these kind of contributions for simplicity. Therefore, for the internal lepton we will take $\ell_m=\tau$,  whereas $V$ will be taken as the  $Z$ gauge boson since we are considering that there are no new particles other than extra scalar bosons. For the three minimal scenarios described above we  show in Fig. \ref{plotneutral} the corresponding contributions to the AWMDM of the tau lepton. For the numerical evaluation we have used the Mathematica numerical routines to evaluate the parametric integrals and a cross-check was done by evaluating the results given in terms of Passarino-Veltman scalar functions with the help of the LoopTools routines \cite{vanOldenborgh:1989wn,Hahn:1998yk}. In the case of type-I contribution we consider two scenarios: a)dominant scalar couplings $S_{i\tau\tau} \gg P_{i\tau\tau}$  and b)dominant pseudoscalar couplings $P_{i\tau\tau}\gg S_{i\tau\tau}$, for $i=1,2$.  Type-II contribution is the only one that develops an imaginary part and we show both its real and imaginary parts. Each contribution is nearly proportional to a product of coupling constants denoted by $C_i$, as indicated in the caption of the Figure.

We observe in the plots of Fig. \ref{plotneutral} that  the tau AWMDM is highly sensitive to an increase in the scalar boson mass and can get suppressed by about one order of magnitude when $m_{\phi_1}$ increases from $200$ to $1000$ GeV.
As far as the real part of $a_\tau^W$ is concerned, in the case of  type-I contribution, a pair of scalar bosons with  scalar couplings larger (smaller) than their pseudoscalar  couplings  give a positive (negative) contribution to the AWMDM, but  in the case of type-II contribution the $CP$-even ($CP$-odd) scalar boson gives a positive (negative) contribution, which means that there could be cancellations between both  contributions.   As far as the type-III contribution is concerned, it is always positive and seems to be slightly larger than type-I and type-II contributions. However  the values shown in the plots of Fig. \ref{plotneutral} could get an additional suppression   when the appropriate values for the coupling constants,  predicted by a specific model, are inserted. We can thus obtain a rough estimate of the tau AWMDM  in a particular model by multiplying the values shown in the plots by the corresponding coupling constants. For instance, if we take either $\|S_{\tau\tau}\|\sim m_\tau/(2m_W) $ or $\|P_{\tau\tau}\|\sim m_\tau/(2m_W)$, type-I and type-II contributions would be suppressed by around four  orders of magnitude with respect to the values shown in the plot, whereas type-III  would be suppressed by  two orders of magnitude. This is due to the fact that type-III contribution involves only one  power of the $S_{\tau\tau}$ or $P_{\tau\tau}$  couplings, whereas  both  type-I and  type-II contributions involve a quadratic power of these couplings,  though type-III contribution can have additional suppression due to the $g_{\phi ZZ}$ coupling. In models with several neutral scalar bosons there could be some enhancement provided that the coupling constants are independent and that there is no cancellation between the distinct contributions. However, sum rules between the coupling constants can prevent that all the scalar couplings can be  simultaneously large. Also, the presence of scalar bosons that are a mixture of $CP$ eigenstates could not be very relevant for the AWMDM as their contribution gives no considerable enhancement, though the most distinctive signature of this scenario would be the appearance of a  WEDM. Finally, the imaginary parts of the type-II contributions to $a_\tau^W$ from  a $CP$-even and $CP$-odd scalar bosons are about the same size but of opposite sign and are one order of magnitude smaller than the real part.

\subsubsection{Singly charged scalar bosons}
We  now turn to focus on the possible contributions from new singly charged scalar bosons, which can give rise to  the three type of contributions   no matter if there is a lone charged scalar boson. Again we will only consider the following minimal scenarios:

\begin{itemize}
\item Type-I contribution   is nonvanishing for a single charged scalar boson, but we will consider the scenarios with both  a single charged scalar  and two nondegenerate charged scalars.
\item Type-II and type-III contributions  require  a single charged scalar boson.
\end{itemize}

\begin{figure}[htb!]
\begin{center}

\includegraphics[width=13cm]{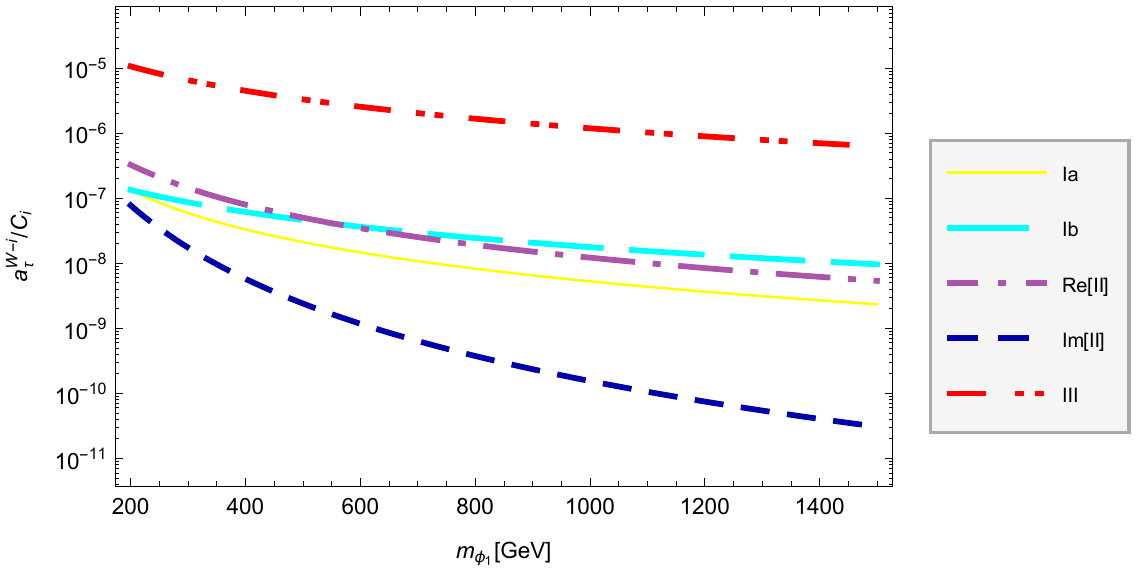}
\caption{Absolute values of the  contributions from charged scalar bosons to the AWMDM of the tau lepton  induced by the three types of  Feynman diagrams of Fig. \ref{FeynmanDiagrams1}. Both the real and imaginary parts of type-II contributions are shown. We consider the following scenarios: two degenerate charged scalar bosons  $ \phi^\pm_1$ and $\phi^\pm_2$ with  $m_{\phi^\pm_2}=m_{\phi^\pm_1}=m_{\phi_1}$ (Ia), two nondegenerate charged scalar bosons  $ \phi^\pm_1$ and $\phi^\pm_2$ with    $m_{\phi^\pm_2}=200$ GeV  (Ib), and a single charged  scalar boson (Re[II], Im[II], and III). We take $P_{i\tau\nu_m}=-S_{i\tau\nu_m}$  for $i=1,2$.
In these scenarios each kind of contribution is nearly proportional to the following product of coupling constants:
$C_{\rm I}=g_{Z\phi^\mp_{1}\phi^\pm_{2}} \|{S_{\tau\nu_\tau}}\|^{2}$,  $C_{\rm II}=\|{S_{\tau\nu_\tau}}\|^{2}$, and $C_{\rm III}=g_{\phi_{1,2}^\mp W^\pm Z}
\textrm{Re}\left[S_{\tau\nu_\tau}\right]$.
\label{charged}}
\end{center}
\end{figure}

The internal lepton $\ell_m$ is now a neutrino $\nu_m$  and $V$ is  the charged $W$  boson.  We will  consider massless neutrinos  so  there would not be lepton-flavor mixing (LFM).   In Fig. \ref{charged} we show the three types of contributions to the AWMDM of the tau lepton in the scenarios described above. For the charged scalar couplings we assume left-handed couplings, namely $P_{i\tau \nu_m}=-S_{i\tau \nu_m}$, $i=1,2$.  Again, the corresponding contributions to the tau AWMDM are nearly proportional to a product of coupling constants, as indicated in the Figure. As for the real parts of $a_\tau^W$, we observe  that  type-I and type-II are now of similar size and   opposite sign, so they can cancel each other out. Thus  type-III contribution is expected to be the dominant one among all the contributions of a singly charged scalar, which is worth noting as the $\phi^\mp W^\pm Z$ vertex is a peculiarity of HTMs. Note also that type-II contribution is the only one that can develop an imaginary part, which is less than one order of magnitude smaller than the real part for $m_{\phi_1}=200$ GeV, but gets considerably suppressed as this mass increases.  Once again, a more careful analysis with appropriate values of the coupling constants is required in a specific model since these  contributions can be considerably suppressed as the coupling constants are expected to be much smaller than the unity.

\subsubsection{Doubly charged scalar bosons}
Doubly charged scalar bosons arise in HTMs and can give contributions to the tau lepton AWMDM from  type-I$'$ and type-II$'$ Feynman diagrams. This class of particles can also arise in models with an extended  gauge sector, which may also predict  doubly charged gauge bosons $Y^{\pm\pm}$. In such a scenario there will be a contribution similar to that arising from type-III diagrams provided that the  $\phi^{\mp\mp}Y^{\pm\pm}Z$ vertex arises at the tree-level. For our analysis we will  consider the same scenarios  as those analyzed in the case of the singly charged scalar boson, though the internal lepton is now a charged one. For simplicity we will assume negligibly LFV couplings  and  take  $\ell_m=\tau$, whose contribution have an extra factor of $4$ due to two vertices with two identical leptons. We will also consider that  the doubly charged  scalar boson couplings are  left-handed, namely,  $P'_{i\tau \tau}=-S'_{i\tau \tau}$, $i=1,2$. Indeed, doubly charged scalar couplings to charged leptons are left-handed (right-handed) if they arise from triplets (doublets). We have analyzed the behavior of the AWMDM of the tau lepton induced by doubly charged scalar bosons in  scenarios I and II of Fig.  \ref{charged}. For similar masses,  the doubly and singly charged scalar contributions only differ by the factor of  4 due to identical leptons. The fact that the internal lepton is now the  tau lepton instead of a massless neutrino does not alter significantly these results. Therefore, the results shown in the curves labeled by Ia, Ib, Re[II] and Im[II] in  Fig.  \ref{charged} are valid for a doubly charged scalar provided that an extra factor of 4 is considered. Also the coupling constants $C_i$ should be replaced  by $C_{\rm I}\to C_{\rm I'}=
g_{Z\phi^{\mp\mp}_{1}\phi^{\pm\pm}_{2}} \|S'_{\tau\tau}\|^2$ and  $C_{\rm II}\to C_{\rm II'}=\|S'_{\tau\tau}\|^2$.
Again, since type-I$'$ and type-II$'$  contributions are of opposite sign, there can be large cancellations between them. If there is no type-III contribution, doubly charged scalar bosons may induce a more suppressed contribution to the tau AWMDM than neutral and singly charged gauge bosons.

\subsection{AWMDM of the tau lepton in models with an extended scalar sector}

We now turn to present an assessment of  the NP contributions to the tau AWMDM from some popular models with an extended scalar sector (we will not consider pure SM contributions). We will only  present a short description of each model as they have been largely studied in the literature. For a review of this class of models see \cite{Ivanov:2017dad} and References therein.

\subsubsection{Singlet models}

These are the simplest extensions of the SM since only add one real or complex singlet $S$  to the SM doublet. Although these models predict  new neutral scalars $h_i$ that may play the role of a dark matter candidate and provide a  connection with  a hypothetical hidden sector (the Higgs portal),  their  phenomenology is not  as interesting as that of models with higher-dimensional multiplets. For instance there are no tree-level flavor change  and new sources of $CP$ violation.   The interactions of the new physical scalars with the SM particles occur via  mixing with the  Higgs doublet since the singlet does not couple to the SM fields.  Therefore,  the new scalar bosons would have suppressed  SM-like couplings to the $Z$ gauge boson and the leptons.  Furthermore, since at least one of the two new neutral scalar bosons would be a dark matter candidate, the tau AWMDM  would receive only  new type-II and type-III contributions  from one of the new scalar bosons at most, whereas type-I would be absent as it only arises when there is $CP$ violation.  We thus conclude that the new contributions  to the tau AWMDM from  singlet models are  not expected to be relevant and we refrain from presenting a more detailed analysis here.

\subsubsection{$CP$ conserving THDMs}

One of the main attractiveness of THDMs is that they are required by supersymmetric theories, but also can have other interesting features, such as  a possible dark matter candidate,  flavor change at the tree-level,  new sources of $CP$ violation,   cosmological implications, etc. After SSB,   the physical Higgs spectrum of $CP$-conserving THDMs is composed by two $CP$-even neutral scalar bosons $h$ and $H$, one  $CP$-odd neutral scalar boson $A$, and a pair of charged scalar bosons $H^\pm$. To forbid tree-level  flavor changing neutral currents (FCNC) a $Z_2$ symmetry is invoked giving rise to four THDMs with natural flavour conservation according to the $Z_2$ charge assignments: model I, model II, lepton-specific model, and flipped model \cite{Branco:2011iw}. On the other hand, the so-called model III is obtained by allowing flavor change in the Yukawa Lagrangian and  constraining  the respective couplings via  experimental data. We do not expect a considerable enhancement of the tau AWMDM if flavor violation is allowed, so we will only consider flavor conserving models.

$CP$-conserving THDMs  would give new type-II contributions to the tau AWMDM  arising from the neutral scalar bosons  $A$ and $H$. The latter would also give a type-III contribution   via the $HZZ$ vertex, though the $g_{hZZ}$ and $g_{HZZ}$ couplings cannot be simultaneously large as the obey the sum rule $g_{hZZ}^2+g_{HZZ}^2=g_{h_{SM}ZZ}^2$.  As for the charged scalar boson $H^\pm$, it would only give  type-I and type-II contributions since the $H^+ W^- Z$ vertex is absent at the tree-level. The nonvanishing contributions have been already studied in the context of model II,  prior to the SM Higgs discovery, \cite{Bernabeu:1995gs} and more recently in model III \cite{Arroyo-Urena:2015uoa}. We will  calculate the results in both model I and model II considering the most up to date bounds on the parameter space. The corresponding Feynman rules have been presented very often in the literature and are summarized in \cite{Branco:2011iw}. In Table \ref{THDMCouplings} we show the  couplings of the scalar bosons necessary for our calculation. From now on we will consider that $h$ is the SM Higgs boson and its couplings have little deviation from  the SM ones.
\begin{table}
\begin{center}
\caption{Nonvanishing couplings of the scalar bosons of  THDMs with natural flavour conservation \cite{Branco:2011iw}. The lepton couplings must be multiplied by $m_\tau/(2m_W)$. Notice that the couplings of the flipped model (lepton-specific model) are the same as those of model I (model II).  \label{THDMCouplings}}
\def\arraystretch{1.2}
\begin{tabular}{ccc}
\hline
\hline
Coupling&Model I& Model II\\
\hline
\hline
$S_{H\tau\tau}$&$\frac{\sin\alpha}{\sin\beta}$&$\frac{\cos\alpha}{\cos\beta}$\\
$P_{A\tau\tau}$&$-i\cot\beta$&$i\tan\beta$\\
$S_{H^-\bar\tau\nu_\tau}$&$\frac{\cot\beta}{\sqrt{2}}$&$-\frac{\tan\beta}{\sqrt{2}}$\\
$P_{H^-\bar\tau\nu_\tau}$&$-S_{H^-\bar\tau\nu_\tau}$&$-S_{H^-\bar\tau\nu_\tau}$\\
$g_{HZZ}$&$\frac{\cos(\alpha-\beta)}{c_W}$&$\frac{\cos(\alpha-\beta)}{c_W}$\\
$g_{ZH^+H^-}$&$\frac{1}{2c_W}(1-2s_{W}^{2})$&$\frac{1}{2c_W}(1-2s_{W}^{2})$\\
\hline
\hline
\end{tabular}
\end{center}
\end{table}

\begin{figure}[htb!]
\begin{center}
\includegraphics[width=17cm]{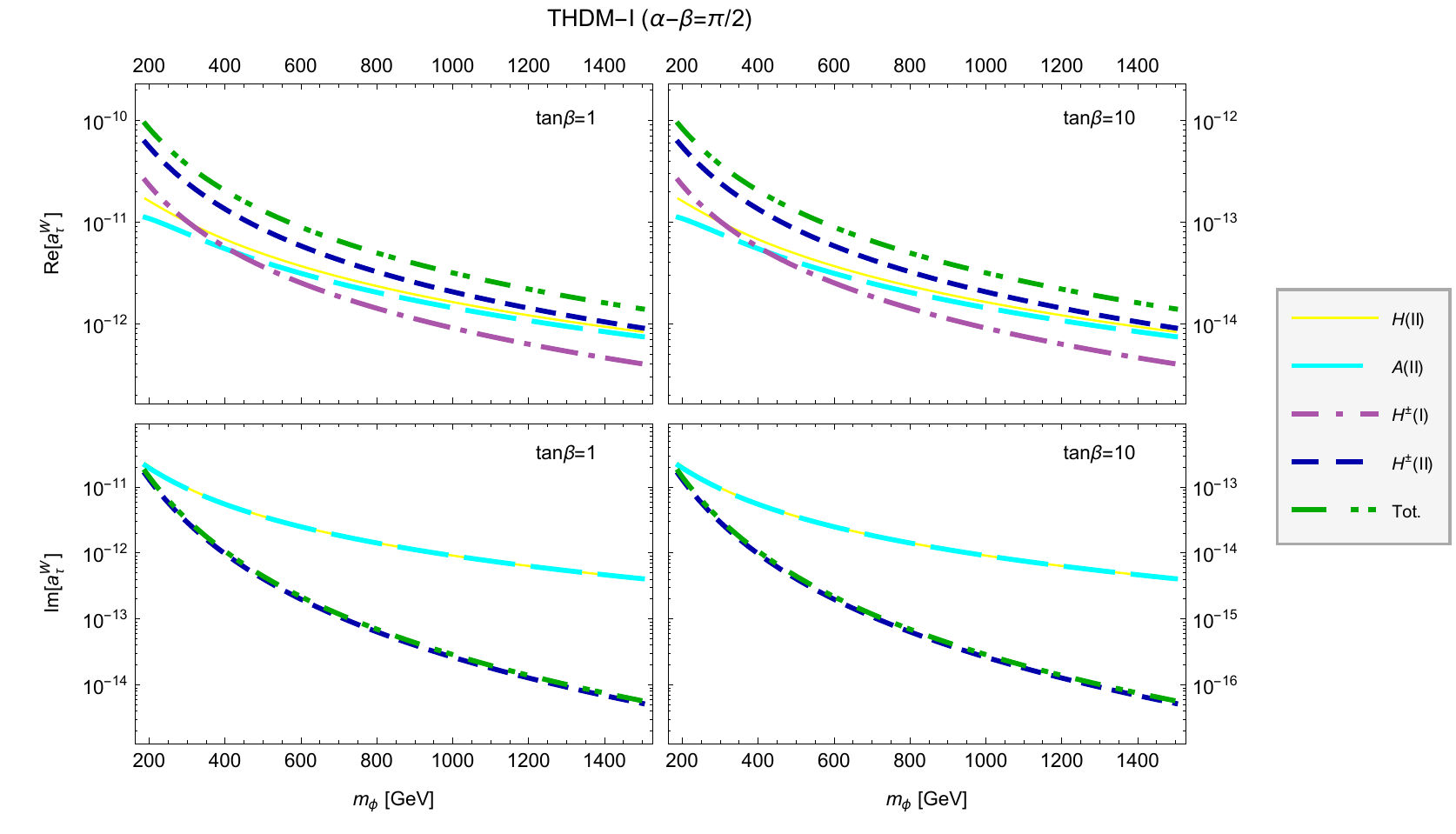}
\caption{Absolute values of the real (upper plots) and imaginary (lower plots) parts of the partial and total contributions from the THDM-I to the AWMDM of the tau lepton as functions of the scalar boson masses considering $m_H=m_A=m_\phi$ and the indicated values of the model parameters.  The contributions of the flipped THDM are identical.\label{plotTHDMI}}
\end{center}
\end{figure}

\begin{figure}[htb!]
\begin{center}
\includegraphics[width=17cm]{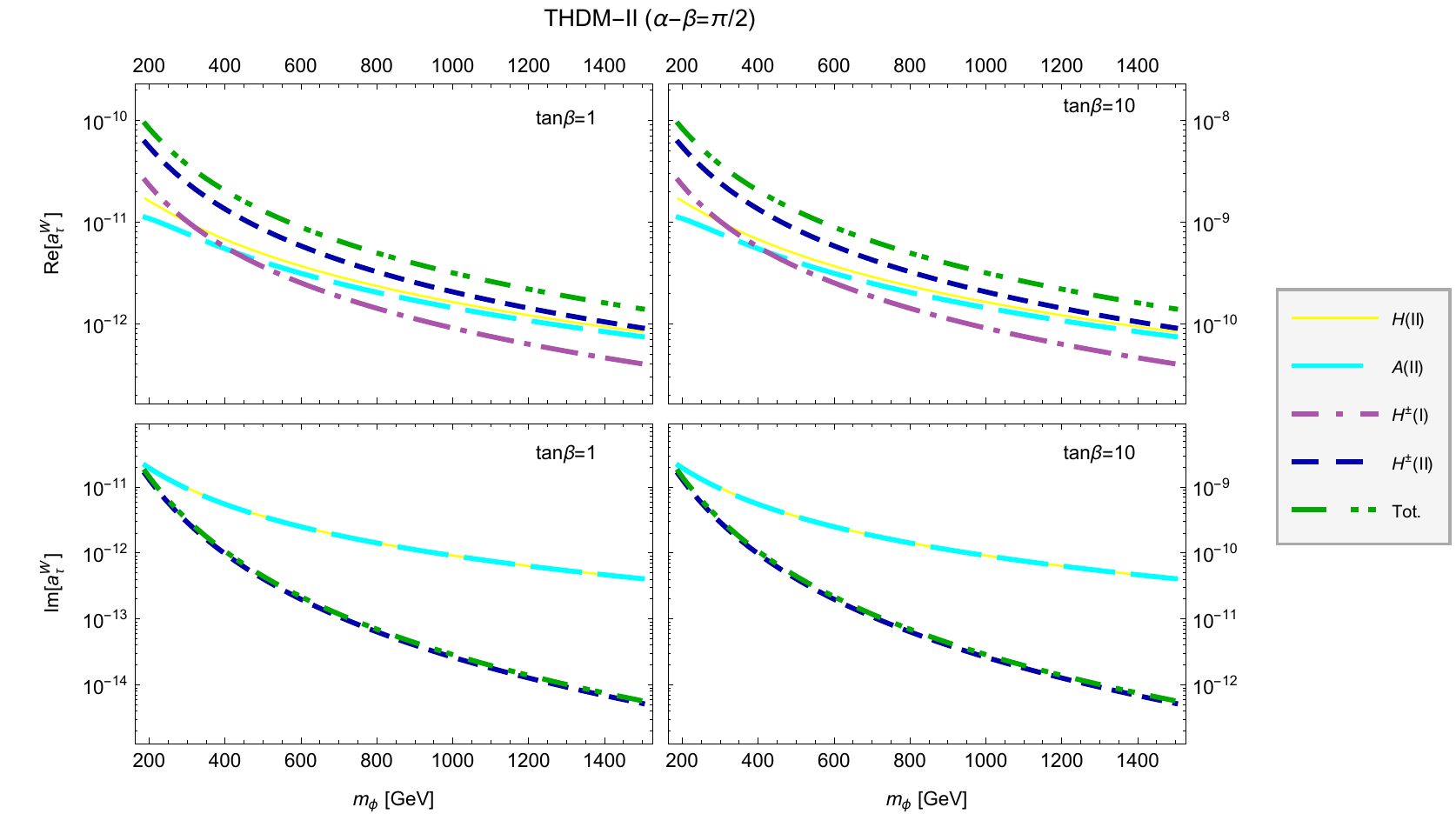}
\caption{The same as in Fig. \ref{plotTHDMI} but for THDM-II. These contributions  are identical to those of the lepton-specific THDM.\label{plotTHDMII}}
\end{center}
\end{figure}

Constraints on the masses of the new scalar bosons and  the parameters $\tan\beta$ and $\alpha$ arise from experimental data. If $h$ is assumed to be the SM Higgs boson, LHC data requires  $\beta-\alpha\simeq \pi/2$ (the alignment limit) and small values of $\tan\beta$. In this scenario, the $HZZ$ vertex would be negligibly small and so would be the corresponding  type-III contribution. In Fig. \ref{plotTHDMI}   we show the behavior of the real and imaginary parts of the partial and total contributions of model I to the tau AWMDM as functions of the scalar boson masses and two values of $\tan\beta$, whereas in Fig. \ref{plotTHDMII} we present the corresponding plots for the contributions of model II. It turns out that such contributions are identical to those of the flipped and the lepton-specific models, respectively. We observe that in model I (model II) all the contribution are proportional to $\cot\beta$ ($\tan\beta$) in the $\beta-\alpha\simeq \pi/2$ limit, thus the total contributions are identical in both models for $\tan\beta=1$, but when this parameter increases its value, $a_\tau^W$ decreases (increases) in model I (model II). The main contributions arise from the charged scalar boson via the type-II diagram, whereas the contributions  of the neutral scalar bosons, both of type-II, are slightly smaller and of opposite sign. Due to the cancellation between the distinct contributions, the total sum of the real part of $a_\tau^{W}$ is of the order of $10^{-10}$ or below for $\tan\beta=1$ and  masses of the scalar bosons above the 200 GeV level. On the other hand, when $\tan\beta=10$, ${\rm Re}\left(a_\tau^{W}\right)$  is of the order of $10^{-12}$  in model I but of order of $10^{-8}$ in model II. These values get considerably suppressed as the scalar boson masses increase.  As far as the imaginary part of $a_\tau^W$ is concerned, both the $H$ and $A$ contributions cancel each other out, so the total contribution is due to the charged scalar boson and is of the order of $10^{-11}$ for   $m_{H^\pm}=200$ GeV, but decreases dramatically  as $m_{H^\pm}$ increases. This contribution is the same in both model I and model II for $\tan\beta=1$, but decreases (increases) by two orders of magnitude when $\tan\beta=10$ in model I (model II). Although we have focused on two values of $\tan\beta$, we can conclude that the contributions of  $CP$-conserving THDMs to the tau AWMDM  are much smaller than the pure SM prediction and are even below the contributions of other SM extensions.

\subsubsection{$CP$ violating THDMs}
The most general THDM allows for $CP$ violation in the Higgs sector, which can arise explicitly (via complex couplings) or spontaneously (via complex VEVs).  We will consider the latter scenario and  follow the approach of the authors of Ref. \cite{Gunion:1997aq,Grzadkowski:1999ye}, where a THDM respecting the $Z_2$ symmetry $\Phi_2\to-\Phi_2$ and $u_{iR}\to -u_{iR}$ is considered (model II). The most general renormalizable Higgs potential that violates softly the $Z_2$ symmetry is given by

\begin{equation}
\label{CPVPotential}
V (\Phi_1,\Phi_2) = V_{sym}(\Phi_1,\Phi_2) + V_{soft}(\Phi_1,\Phi_2),
\end{equation}
where the $Z_2$ symmetric term is
\begin{eqnarray}
V_{sym}(\Phi_1,\Phi_2) &=& -\mu^2_{1}\Phi^\dagger_1\Phi_1 -\mu^2_{2}\Phi^\dagger_2\Phi_2
+\lambda_1(\Phi^\dagger_1\Phi_1)^2 + \lambda_2(\Phi^\dagger_2\Phi_2)^2+ \lambda_3(\Phi^\dagger_1\Phi_1)(\Phi^\dagger_2\Phi_2)
+\lambda_4|\Phi^\dagger_1\Phi_2|^2\nonumber\\
&+&\frac{1}{2}\left( \lambda_5(\Phi^\dagger_1\Phi_2)^2 + {\rm H.c.}\right),
\end{eqnarray}
whereas the softly-violating term is given by
\begin{equation}
V_{soft}(\Phi_1,\Phi_2)=-\mu^2_{12}\Phi^\dagger_1\Phi_2+{\rm H.c.}
\end{equation}

After SSB  $\Phi_1$ and $\Phi_2$ acquire the VEVs $\Phi_1=\langle v_1 \rangle/\sqrt{2}$ and  $\Phi_2^T=\langle v_2 e^{i\theta} \rangle/\sqrt{2}$ as long as $|\mu^2_{12}/(2\lambda_5v_1v_2)|< 1$. The three neutral physical states $h_i$, which are now a mixture of $CP$-even and  $CP$-odd eigenstates, are obtained from the gauge eigenstates as follows   $h_i=R_{ij}\phi_j$, with $h_i^T=(h_1,h_2,h_3)$, $\phi^T=\sqrt{2}({\rm Re}\left(\phi_1^0\right),{\rm Re}\left(\phi_2^0\right), \left(s_\beta {\rm Im}\left(\phi^0_1\right)-c_\beta{\rm Im}\left(\phi^0_2\right)\right))$ and $R=R_3 R_2 R_1$  a rotation matrix that can be parametrized as
\begin{equation}
\label{RotationMatrix}
{\bf R}=\begin{pmatrix}
c_1 &-s_1c_2 &s_1s_2\\
s_1c_3 &c_1c_2c_3-s_2s_3&-c_1 s_2c_3-c_2s_3\\
s_1s_3& c_1c_2s_3+s_2c_3&-c_1s_2s_3+c_2c_3
\end{pmatrix},
\end{equation}
where $s_i\equiv\sin\alpha_i$ and $c_i\equiv\cos\alpha_i$, with $\alpha_i$ the Euler angles ($i=1,2,3$). The $CP$-conserving THDM-II is obtained in the limit of $\alpha_2=\alpha_3=0$, after which one must redefine $\alpha_1=\pi/2-\alpha$ to get the conventional nomenclature.

In the Yukawa sector, after SSB the $\ell_l \ell_l h_i$ couplings  acquire the form of Eq. (\ref{Hff}). As far as the scalar-to-gauge-bosons interactions are concerned, there are not only $h_i ZZ$ couplings but also $Z h_i h_j$ ones ($i,j=1,2,3$, $i\ne j$), whereas the charged gauge couplings are the same as those of the $CP$-conserving THDM-II. The Feynman rules for this model are presented in \cite{Gunion:1997aq}.  In Table \ref{CPVTHDMCouplings} we show a summary of the coupling constants required for the evaluation of the tau AWMDM.

\begin{table}
\begin{center}
\caption{Nonvanishing couplings of the scalar bosons in $CP$-violating  THDM-II \cite{Gunion:1997aq}. The lepton couplings must be multiplied by $m_\tau/(2m_W)$. The $C_i$ and $C_{ij}$ constants are $C_i=s_\beta R_{i2}+c_\beta R_{i1}$ and  $C_{ij}=w_i R_{j3}-w_j R_{i3}$, with  $w_i=s_\beta R_{i1}-c_\beta R_{i2}$. Note that the couplings of charged scalars remain unchanged. \label{CPVTHDMCouplings}}
\def\arraystretch{1.2}
\begin{tabular}{cc}
\hline
\hline
Vertex&Coupling constant\\
\hline
\hline
$S_{h_i\tau\tau}$&$\frac{R_{i1}}{\cos\beta}$\\
$P_{h_i\tau\tau}$&$-i\tan\beta R_{i3}$\\
$S_{H^-\bar\tau\nu_\tau}$&$\frac{\tan\beta}{\sqrt{2}}$\\
$P_{H^-\bar\tau\nu_\tau}$&-$S_{H^-\bar\tau\nu_\tau}$\\
$g_{h_iZZ}$&$\frac{C_i}{c_W}$\\
$g_{Z h_i h_j}$&$\frac{C_{ij}}{2c_W}$\\
$g_{ZH^+H^-}$&$\frac{1}{2c_W}(1-2s_{W}^{2})$\\
\hline
\hline
\end{tabular}
\end{center}
\end{table}

\begin{figure}[htb!]
\begin{center}
\includegraphics[width=17cm]{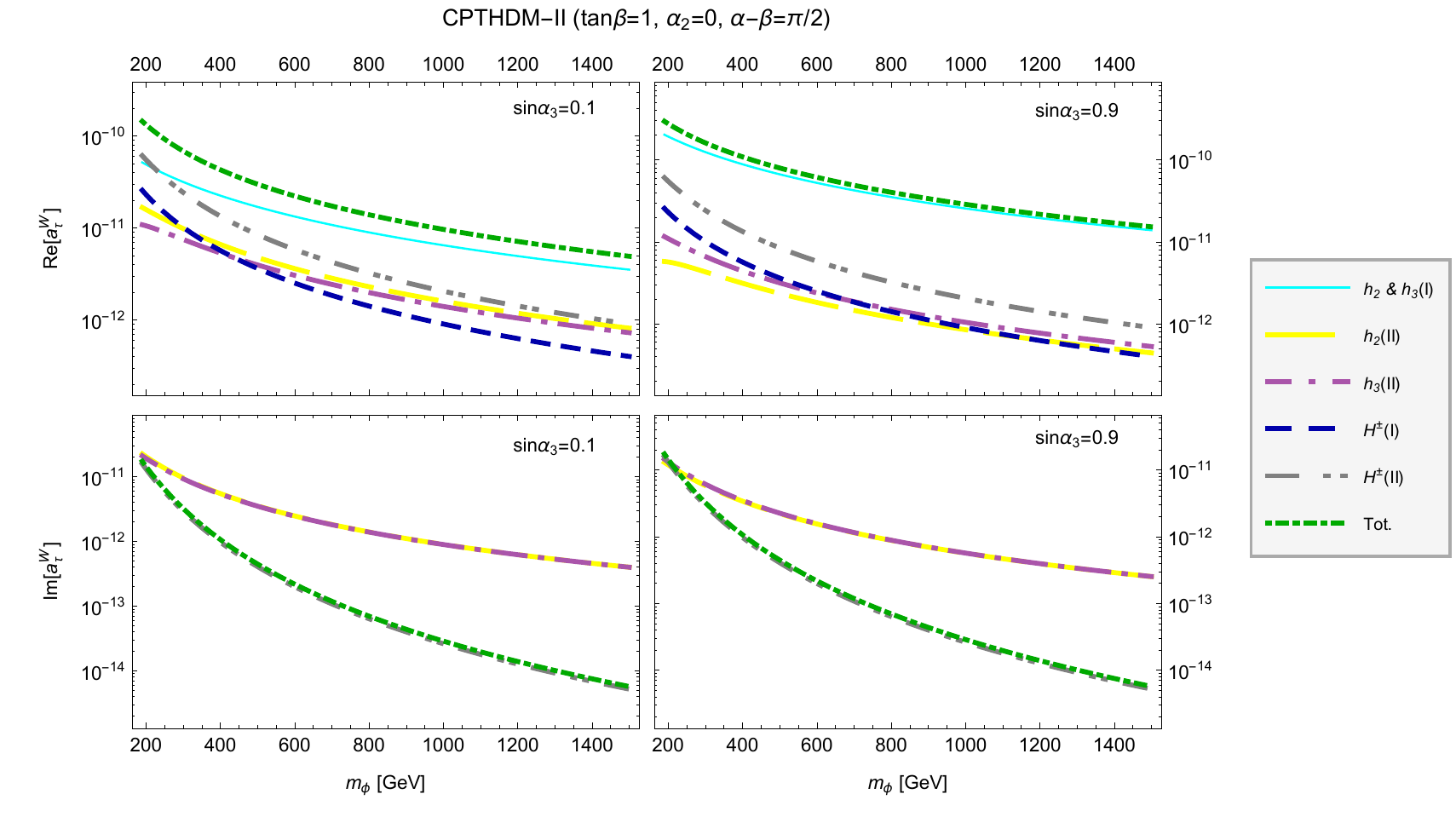}
\caption{Absolute values of the real (upper plots) and imaginary (lower plots) parts of the partial and total contributions from the $CP$-violating THDM-II to the AWMDM of the tau lepton as functions of the scalar boson masses considering $m_{h_1}=200$ GeV and $m_{h_3}=m_{\phi_1}$  and the indicated values of the model parameters.  \label{plotCPTHDM}}
\end{center}
\end{figure}

In the most general scenario, the tau AWMDM  receives contributions from the three neutral scalars   via the three types of Feynman diagrams of Fig. \ref{FeynmanDiagrams1}, whereas the charged scalar boson would contribute through   type-I and type-II diagrams. However, it must be noted that the following sum rule is obeyed by the couplings of the scalar bosons to the $Z$ gauge boson \cite{Gunion:1997aq}: $C_i^2+C_j^2+C_{ij}^2=1$ ($i\ne j$), which means that even though there are additional contributions, some of them would be negligible since not all the coupling constants can increase simultaneously. The properties of the 125 GeV Higgs boson discovered at the LHC seem to fit very well with those of the SM Higgs boson,  we will thus consider  that  $h_1$ is the lightest scalar boson and its properties are nearly identical to the 125 GeV SM Higgs boson, namely $C_1\simeq 1$. This scenario  corresponds to $\alpha_2\simeq0$, therefore $h_1$ becomes a pure $CP$ even state and the coupling constants become in this limit $C_1\sim C_{23}\sim \sin(\alpha-\beta)\simeq 1$, whereas all of $C_2$, $C_3$, $C_{12}$, and $C_{13}$ become proportional to $\cos(\alpha-\beta)\simeq 0$. The resulting coupling constants are shown in Table \ref{CPVTHDMWEDMCouplings}, where we have neglected all terms proportional to $\cos(\alpha-\beta)$. In this case  there would only be a new type-I contribution arising from the scalar bosons $h_2$ and $h_3$, which also would contribute through the type-II  diagrams but not the type-III one. In addition, there are also  contributions of the charged scalar boson, which are the same as in the $CP$-conserving THDM-II as the charged scalar couplings remain unchanged. In Fig. \ref{plotCPTHDM} we show the partial and total contributions to the tau AWMDM for $\tan\beta=1$ and two values of $\sin\alpha_{3}$. We observe that the real part of $a_\tau^W$ is now dominated by the type-I contribution but it is of the same order of magnitude than the contribution of the $CP$-conserving THDM-II. On the other hand, the imaginary part of $a_\tau^W$ is dominated by the contributions of the charged scalar boson since the contributions from the neutral scalar bosons cancel each other out. Thus the imaginary part of $a_\tau^W$ is very similar to that of the $CP$-conserving model. We thus conclude that the contributions of the $CP$-violating THDM give no relevant enhancement to the tau AWMDM as compared to the $CP$-conserving model.  However, the most interesting implication of this scenario is the appearance of a WEDM as discussed below.

\begin{table}
\begin{center}
\caption{Nonvanishing couplings of the scalar bosons in  the $CP$-violating  THDM-II in the scenario with $\alpha_2=0$ and $\cos(\alpha-\beta)\simeq 0$. We set $\alpha_1=\pi/2-\alpha$. The lepton couplings must be multiplied by $m_\tau/(2m_W)$. \label{CPVTHDMWEDMCouplings}}
\def\arraystretch{1.2}
\begin{tabular}{cc}
\hline
\hline
Vertex&Coupling constant\\
\hline
\hline
$S_{h_2\tau\tau}$&$-\frac{\cos\alpha\cos\alpha_3}{\cos\beta}\simeq \tan\beta\cos\alpha_3$\\
$P_{h_2\tau\tau}$&$i\tan\beta \sin\alpha_3$\\
$S_{h_3\tau\tau}$&$-\frac{\cos\alpha\sin\alpha_3}{\cos\beta}\simeq \tan\beta\sin\alpha_3$\\
$P_{h_3\tau\tau}$&$-i\tan\beta \cos\alpha_3$\\
$g_{Z h_2 h_3}$&$-\frac{\sin(\alpha-\beta)}{2c_W}$\\
\hline
\hline
\end{tabular}
\end{center}
\end{table}

\subsubsection{Multiple Higgs doublet models}
Since the $\rho=1$ relation remains valid  at the tree level after  the addition of an arbitrary number of Higgs doublets to the SM scalar sector,  models with more than two Higgs doublets  have  also been the focus of considerable attention in the literature, though they are plagued with a large number of free parameters \cite{Grossman:1994jb}.  Apart from their rich phenomenology, MHDMs are particularly appealing as they allow for a plethora of discrete and Abelian symmetries in the scalar and flavor sectors \cite{Ivanov:2011ae,Ivanov:2012ry,Ivanov:2014doa}, which can be useful to reduce the number of free parameters.
An $N$-Higgs doublet model predicts $N-1$ pairs of new charged scalar bosons and $2N-1$ neutral scalar bosons,  including the SM Higgs boson. If $CP$ conservation is assumed in the scalar sector, the neutral scalar bosons are $CP$ eigenstates, but this is not true when $CP$ violation is allowed. Thus, in principle there would be more extra contributions to the tau AWMDM. However, a phenomenologically viable MHDM requires that one of the neutral scalar boson has couplings to the  SM gauge bosons and fermions very similar to those of the SM Higgs boson. Therefore, it is often assumed that all the neutral Higgs bosons other than the SM-like one are heavy and have suppressed couplings to the SM particles (the decoupling limit). As far as the charged scalar bosons are concerned, their couplings to lepton pairs cannot be simultaneously large due to the corresponding sum rules. It has been pointed out that the more interesting  scenario  is that in which all the extra charged scalar bosons but only one  decouple from the fermions \cite{Grossman:1994jb}. Therefore, we do not expect that the extra contributions of a MHDM show an important enhancement to the tau AWMDM compared to the contribution of a THDM since the extra neutral and charged scalar bosons could be very heavy and  have suppressed couplings to the tau lepton and the $Z$ gauge boson.

\subsubsection{The Georgi-Machacek Model}
Although this class of models can give dangerous contributions to the $\rho$ parameter, this can be alleviated if  either a custodial $SU(2)_C$ symmetry is imposed \cite{Georgi:1985nv,Chanowitz:1985ug} or the Higgs triplets VEVs are  of the order of a few GeVs \cite{Schechter:1980gr}. The former realization of Higgs triplet models is known as the Georgi-Machacek model (GMM) and the latter  as the Higgs triplet model (HTM).  The most interesting features of these models are the following:   naturally light Majorana masses for the neutrinos  via the so called type-II see-saw mechanism,   enhanced Higgs-to-gauge boson couplings,     doubly charged scalar bosons,  and tree-level induced $H^\pm W^\mp Z$ coupling  \cite{Gunion:2006ga}. The last two features are a unique signature of these models, which can offer a  rich phenomenology and  provide a clear signal at particle colliders.  For instance, doubly charge scalars  can enhance significantly  the two-photon decay of a neutral Higgs boson.

We will first discuss the scenario posed by the GMM, which predicts a more rich physical spectrum. This model contains the usual SM doublet $\phi$, one real triplet $\Xi=(\xi^+, \xi^0, \xi^-)^T$ with $Y=0$ and one complex triplet $X=(\chi^{++}, \chi^{+}, \chi^0)^T$ with $Y=2$. They are arranged in a bidoublet

\begin{equation}
\Phi=\left(\begin{array}{cc}
\phi^{0*} & \phi^{+}\\
-\phi^{+*} & \phi^{0}
\end{array}\right),
\end{equation}
and a bitriplet
\begin{equation}
\Delta(\tilde X, \Xi,X)=\left(\begin{array}{ccc}
\chi^{0*} & \xi^{+} & \chi^{++}\\
-\chi^{+*} & \xi^{0} & \chi^{+}\\
\chi^{++*} & -\xi^{+*} & \chi^{0}
\end{array}\right),\label{fields}
\end{equation}
where $\Phi$ and $X$ transform under the custodial symmetry as $\Phi\to U_L\Phi U_R^{\dagger}$ and  $\Delta \to U_L \Delta U_R^{\dagger}$ with $U_{L,R}=e^{(i\theta^a_{L,R}T^a)}$. Here $T^a=t^a$ stands for the $SU(2)$ generators in the triplet representation. To achieve SSB, the neutral components of the doublet and the triplets acquire VEVs $v_\phi$, $v_\xi$ and $v_\chi$. It turns out that $\rho=1$ at the tree-level since $v_\xi$ and  $v_\chi$ are aligned due to the custodial symmetry. After SSB, nine physical scalar bosons emerge: apart from  the SM-like Higgs boson $h$, there are  one scalar singlet $H$, one scalar triplet $H_3$ ($H_3^0$, $H_3^\pm$), and one scalar fiveplet $H_5$ ($H_5^0$, $H_5^{\pm\pm}$, $H_5^{\pm}$). While $H$ and $H_5^0$ are $CP$-even, $H_3^0$ is $CP$-odd. A peculiarity  of this model is that the triplet and fiveplet masses are degenerate. Also, since there is no doublet field in the custodial fiveplet,  the $H_5$   states are fermiophobic and their couplings to a fermion antifermion pair can only arise via  radiative corrections.  The Feynman rules for the GMM are presented in \cite{Chiang:2012cn}. In Table \ref{GMMcouplings} we show the  couplings necessary to evaluate  the tau AWMDM.

\begin{center}
\begin{table}[!htb]
\caption{Coupling constants for vertices inducing the tau AWMDM in the GMM \cite{Chiang:2012cn}. Here $\theta_H$ and $\alpha$ are mixing angles: $\sin \theta_H=\sqrt{2}2v_\xi/v$ and $\alpha$ is the mixing angle of the $h$ and $H$ scalar bosons.  The lepton couplings must be multiplied by $m_\tau/(2m_W)$. Although the   ${ZH_{3}^{0}h}$,
${ZH_{3}^{0}H}$, and
${Z H_{3}^{0}H_{5}^{0}}$ couplings also arise, as explained in the text they do not contribute to the tau AWMDM at the one-loop level.
\label{GMMcouplings}}
\begin{tabular}{cc}
\hline
\hline
Vertex & Coupling constant\\
\hline
\hline
$S_{H\tau\tau}$&$\frac{\sin\alpha}{\cos\theta_H}$\\
$P_{H_3^0\tau\tau}$&$i\tan\theta_H$\\
$S_{H_3^-\bar\tau\nu_\tau}$&$\frac{\tan\theta_H}{\sqrt{2}}$\\
$P_{H_3^-\bar\tau\nu_\tau}$&-$P_{H_3^-\bar\tau\nu_\tau}$\\
$g_{ZH_{3}^{+}H_{3}^{-}}$ & $\frac{1}{2c_{W}}(1-2s_{W}^{2})$\\
$g_{H ZZ}$ & $\frac{1}{3c_W}(3c_H s_\alpha-2\sqrt{6} s_H c_\alpha)$\\
\hline
\hline
\end{tabular}
\end{table}
\end{center}

\begin{figure}[htb!]
\begin{center}
\includegraphics[width=17cm]{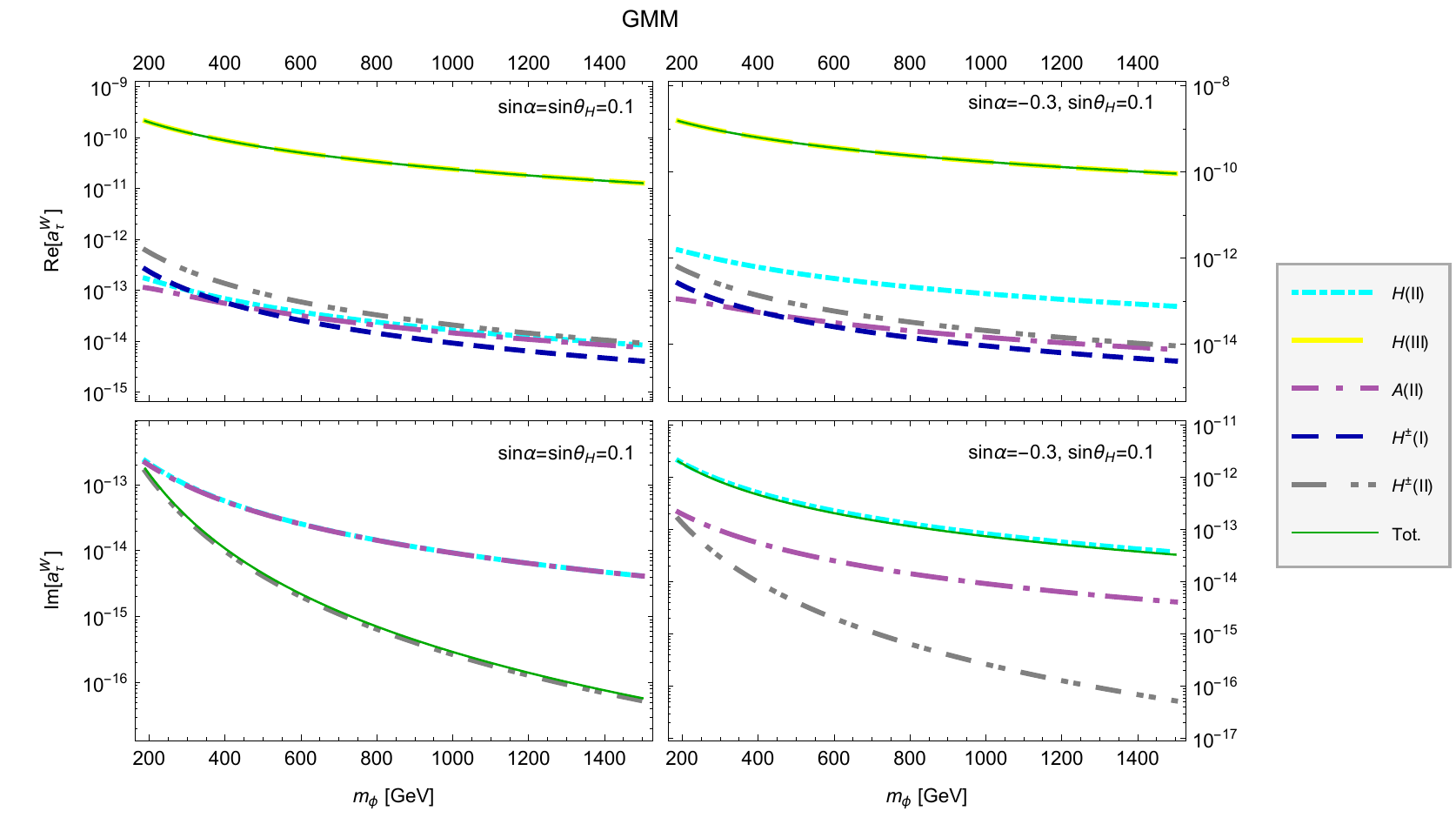}
\caption{Absolute values of the real (upper plots) and imaginary (lower plots) parts of the partial and total contributions from the GMM to the AWMDM of the tau lepton as functions of the scalar boson masses considering $m_H=m_{H^0_3}=m_\phi$ and the indicated values of the model parameters. \label{plotGMM}}
\end{center}
\end{figure}

In spite of the  wide spectrum of physical scalar bosons of the GMM, there is only a handful of new contributions to $a_\tau^W$: the $H$ scalar boson would give contributions of type-II and type-III, $H_3^0$ would only induce a type-II contribution, and $H_3^\pm$ would contribute via the type-I and type-II diagrams. It is worth noting that  since the fiveplet is fermiophobic, there is no type-III contribution from neither the singly charged scalar $H^{\pm}_5$ (via the $H^\pm_5 W^\mp Z$ coupling) nor  the doubly charged scalar $H^{\pm\pm}_5$. Therefore, the contribution of the GMM to would be similar to that of a THDM. We consider two set of values for the $\sin\alpha$ and $\sin\theta_H$ parameters still consistent with the constraints on the parameter space of the GMM \cite{Hartling:2014aga,Chiang:2014bia} and plot the behavior of the partial and total contributions to the tau AWMDM as functions of the scalar boson masses. The results are shown in Fig. \ref{plotGMM}, where it is observed that the real part of the type-III contribution arising from the $H$ scalar boson is the dominant one, whereas the real parts of the remaining contributions are considerably suppressed. As for the $a_\tau^W$ imaginary part, for $\sin\alpha=\sin\theta_H=0.1$ the contributions from the $H$ and $H^0_3$ scalar bosons are nearly identical but of opposite sings, therefore they cancel out and the bulk of the imaginary part of $a_\tau^W$ is due to the type-II contribution from the charged scalar boson $H^\pm_3$. On the other hand, when $\sin\alpha=-0.3$ and $\sin\theta_H=0.1$ the dominant contribution arises from the $H$ scalar boson, whereas the remaining contributions are negligible. In general, the real part of the tau AWMDM can reach the level of  $10^{-9}-10^{-10}$ for masses of the new scalar bosons of the order of 200 GeV, whereas the imaginary part is of the order of $10^{-12}-10^{-13}$. We observe that the behavior of $a_\tau^W$ is highly sensitive to the values of $\sin\alpha$ and $\sin\theta_H$. It is worth noting that in this model there is no sum rule for the $g_{hZZ}$ and $g_{HZZ}$ couplings thus the type-III contribution from the $H$ scalar boson can be relevant for the total contribution. We can conclude that  although in the GMM there is a slight enhancement of $a_\tau^W$  as compared to THDMs with $\tan\beta=1$ and $\alpha-\beta\simeq \pi/2$, the results are still below the contributions of other SM extensions.

\subsubsection{The Higgs-triplet model}
The HTM is another  comprehensively studied realization of triplet models \cite{Schechter:1980gr}. The main aim of this model is to generate small neutrino masses. In such a model only one complex triplet $\Delta$ with $Y=1$ is introduced along with the SM doublet $\Phi$. The triplet is arranged as

\begin{equation}
\Delta=\left(\begin{array}{cc}
\frac{1}{\sqrt 2}\delta^{+} & \delta^{++}\\
\delta^{0} & -\frac{1}{\sqrt 2}\delta^{+}
\end{array}\right).\label{TripletHTM}
\end{equation}
The neutrino masses are generated via the following Yukawa Lagrangian

\begin{equation}\label{YukawaHTM}
{\cal L}_Y=-{Y_\nu}_{ij} L_{i}^T C i\sigma_2 \Delta L_{j}+{\rm H.c.},
\end{equation}
with $L^T_{i}=(\nu^T_{iL}, e^T_{iL})$ and ${\bf Y}_\nu$ a symmetric complex matrix.

The Higgs doublet  and  the neutral component of $\Delta$ acquire  VEVs $v_0$ and $v_\Delta$, with $\sqrt{v_0^2+v_\Delta^2}=v=246$ GeV. To satisfy the $\rho\simeq 1$ constraint the triplet VEV must fulfill $1$ eV$\leq v_\Delta \le 8$ GeV. After SSB,   seven physical scalar bosons are left as  remnant: two $CP$-even scalar bosons $H_1$ and $H_2$, one $CP$-odd scalar boson $A$, a pair of charged scalar bosons $H^\pm$, and a pair of doubly charged scalar bosons $H^{\pm\pm}$. Other than the doubly charged scalar $H^{\pm\pm}=\delta^{\pm\pm}$, the mass eigenstates are mixtures of the doublet $\phi$ and the triplet $\Delta$. The mixing  is proportional to $v_\Delta$, thus $H_2$, $A$ and $H^\pm$ are mainly composed of the triplet fields, whereas  $H_1$  is predominantly composed of the doublet field and can be identified with the SM Higgs boson. All the extra scalar bosons are nearly mass degenerate. From \eqref{YukawaHTM},  the following neutrino mass matrix is obtained

\begin{equation}
\label{HTNeutrinoMasss}
{\bf M}_\nu=\sqrt{2}v_\Delta{\bf Y}_\nu=\sqrt{2}{\bf V}^*_{PMNS}{\bf M}_\nu^{diag} {\bf V}^\dagger_{PMNS},
\end{equation}
with ${\bf V}_{PMNS}$ the Pontecorvo–Maki–Nakagawa–Sakata mixing matrix.

A more detailed description of this model and the corresponding Feynman rules are presented in Ref. \cite{Perez:2008ha}. The coupling constants necessary for the evaluation of $a_\tau^W$ in the HTM are shown in Table \ref{HTMcouplings}. In particular, the couplings of the charged and doubly charged scalar bosons to leptons are
\begin{equation}
\label{Gamma+}
{\bf \Gamma}_+=\cos\theta_+ \frac{{\bf M}_\nu^{diag} {\bf V}^\dagger_{PMNS}}{v_\Delta}
\end{equation}
and
\begin{equation}
\label{Gamma++}
{\bf \Gamma}_{++}=\frac{{\bf M}_\nu}{\sqrt{2}v_\Delta}.
\end{equation}

Apart from the  $H_1$ contribution to $a_\tau^W$, which corresponds to the SM, the new contributions  to the  tau AWMDM arise from the neutral scalar bosons $H_2$ and $A$ as well as the singly and doubly charged scalar bosons $H^\pm$ and $H^{\pm\pm}$. Contrary to the case of the GMM,  the charged scalar boson now does yield a type-III contribution via the $H^\pm W^\mp Z$ vertex.  Therefore, as compared to THDMs,  the tau AWMDM has the following additional contributions:
a type-III contribution from the singly charged scalar boson and type-I and type-II contributions from the doubly charged scalar bosons. However, all these contributions are highly dependent on the value of $v_\Delta$ and can be considerably suppressed. Let us first examine the contributions of type I and II arising from the charged and doubly charged scalar bosons. Since the neutrino mass matrix elements are typically of the order of  $10^{-2}-10^{-3}$ eV \cite{Olive:2016xmw}, for  $v_\Delta$ of the order of $10^{-5}$ GeV, the matrix elements $\Gamma_+^{ij}$ and $\Gamma_{++}^{ij}$ would be of the order of  $10^{-6}-10^{-7}$ and even smaller for larger values of $v_\Delta$, which means that  type-I and type-III contributions arising from the charged and doubly charged scalar bosons would be negligibly small, below the $10^{-15}$ level. As far as type-II contributions from the neutral scalar boson are concerned, current constraints from unitarity, the oblique $T$ parameter and the diphoton strength signal  \cite{Das:2016bir} favor the regime where the mixing angles are small and can be approximated as $\sin\theta_0\simeq \sin \alpha\simeq 2v_\Delta/v_0$ and $\sin\theta_+\simeq \sqrt{2}v_\Delta/v_0$, so the contributions to the tau AWMDM from the neutral scalar bosons are proportional to $(v_\Delta/v_0)^2$ and so are expected to be very suppressed: for $v_\Delta$ around 1 GeV $v_\Delta/v_0\simeq 8\times 10^{-3}$, thus the $H_2$ and $A$ contributions would be of the order of $10^{-15}$.  A similar result is true for type-III contributions arising from the scalar bosons $H_2$ and  $H^\pm$ since all of the $H_2 ZZ$ and $H^\pm W^\mp Z$ coupling constants are proportional to $v_\Delta/v_0$, thus the respective contribution to $a_\tau^W$ would be proportional to $(v_\Delta/v_0)^2$.  In conclusion, the extra contributions from the HTM to the tau AWMDM would be much smaller than those of THDMs and the GMM.

\begin{center}
\begin{table}[!htb]
\caption{Coupling constants for the vertices inducing the tau AWMDM in the HTM \cite{Perez:2008ha}.   ${\bf \Gamma}_+$ and ${\bf \Gamma}_{++}$ are given in \eqref{Gamma+} and \eqref{Gamma++}. The $S_{H_2\tau\tau}$ and $P_{A\tau\tau}$ couplings must be multiplied by $m\tau/(2m_W)$. We used the approximation $v_0\simeq v$.
\label{HTMcouplings}}
\begin{tabular}{cc}
\hline
\hline
Vertex & Coupling constant\\
\hline
\hline
$S_{H_2\tau\tau}$&$-\sin\theta_0$\\
$P_{A\tau\tau}$&$-i\sin\alpha$\\
$S_{H^-\bar\ell_i\nu_j}$&$\frac{1}{g}\Gamma^{ij}_+$\\
$P_{H^-\bar\ell_i\nu_j}$&-$S_{H^-\bar\ell_i\nu_j}$\\
$S_{H^{--}\ell_i\ell_j}$&$\frac{1}{g}\Gamma^{ij}_{++}$\\
$P_{H^{--}\ell_i\ell_j}$&-$S_{H^{--}\ell_i\ell_j}$\\
$g_{H^+W^-Z}$ & $-s_W t_W\left(\sin\theta_+ -\sqrt{2}\left(2+\frac{1}{t_W^2}\right) \cos\theta_+\frac{v_\Delta}{v_0}\right)$\\
$g_{H_2 ZZ}$ & $\sin\theta_0-4\cos\theta_0 \frac{v_\Delta}{v_0}$\\
$g_{ZH^{+}H^{-}}$ & $\frac{1}{2c_{W}}\left(1-2s_{W}^{2}\right)$\\
$g_{ZH^{++}H^{--}}$ & $\frac{1}{c_{W}}\left(1-2s_{W}^{2}\right)$\\
\hline
\hline
\end{tabular}
\end{table}
\end{center}

\subsubsection{Models with exotic scalar sectors}

The idea of invoking a $SU(2)$ custodial symmetry to preserve the $\rho=1$ relation at the tree level can be generalized to higher-dimensional multiplets, which is achieved by replacing the Higgs bitriplet by a larger representation under the $SU(2)_L\times SU(2)_R$ symmetry \cite{Logan:2015xpa}. These models, dubbed generalized GMM, have a spectrum of physical scalar bosons composed by the singlet $H$, the triplet $H_3$, plus higher  fermiophobic multiplets (a fiveplet, a septet, etc.)  Therefore, their contribution to the tau AWMDM would be similar to that of the  minimal GMM.

Other models with an exotic scalar sector can be constructed by adding extra higher-dimensional multiplets that respect the relation $\rho=1$   at the tree level. Among such class of multiplets,  the  lowest-dimensional  is a septet with $T=3$ and $Y=4$. A model of this class has been investigated quite recently \cite{Alvarado:2014jva} and it has been dubbed the doublet-septet model. However, the septet does not couple directly to the SM fermions and its interactions only arises  through the mixing with the SM doublet. Fifteen physical scalar bosons emerge after SSB, but the relevant ones for the tau AWMDM are two neutral $CP$-even scalars $h$ and $H$, one neutral $CP$-odd scalar $A$, and two pairs of charged scalars $H_1^\pm$ and $H_2^\pm$. Among the remaining physical fields there are a fermiophobic doubly charged scalar and higher-charged scalars. Therefore, the new contribution of this model as compared to that of a THDM arises from the extra charged scalar. Thus, we also do not expect a substantial increment to the tau AWMDM from the doublet-septet model.

\subsection{WEDM of the tau lepton}
\begin{figure}[htb!]
\begin{center}
\includegraphics[width=13cm]{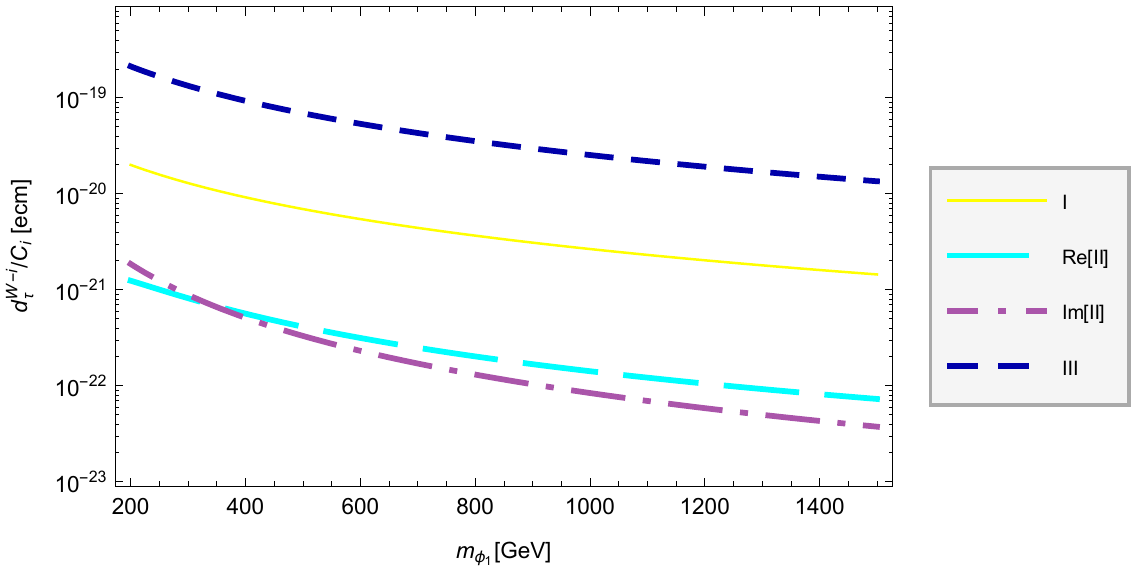}
\caption{Absolute values of the  contributions from new neutral scalar bosons  to the WEDM of the tau lepton  induced by the three types of  Feynman diagrams of Fig. \ref{FeynmanDiagrams1}. Both the real and imaginary parts of type-II contributions are shown. We consider the following scenarios: two nondegenerate scalar bosons  $\hat \phi^0_1$ and $\hat \phi^0_2$ with  $m_{\hat\phi^0_1}=m_{\phi_1}$  and  $m_{\hat\phi^0_2}=200$ GeV (I) as well as a single   scalar boson $\hat\phi^0_1$ (Re[II], Im[II], and III). For  type-I contribution  we take for simplicity $S_{1\tau\tau}P^*_{2\tau\tau}\simeq S_{2\tau\tau} P^*_{1\tau\tau}$, whereas for type-III contribution we use $S_{1\tau\tau} >>P_{1\tau\tau}$. In these scenarios, each  kind of contribution is proportional to the following product of coupling constants:
$C_{\rm I}=g_{Z\phi_1\phi_2} \textrm{Im}\left[S_{1\tau\tau}P_{2\tau\tau}^*\right]$,  $C_{\rm II}={\rm Im}\left[{S_{1\tau\tau}}P_{1\tau\tau}^*\right]$,   and   $C_{\rm III}= g_{\phi_{1}ZZ}
\textrm{Im}\left[S_{1\tau\tau}\; \right]$.  \label{plotWEDMneutral}}
\end{center}
\end{figure}

We now turn to analyze the contributions of new scalar bosons to the tau  WEDM, which requires a $CP-$violating phase. Following along the same line of discussion as in the AWMDM, we will only focus on those models with an extended scalar  sector, therefore we will not consider additional gauge bosons or fermions. However, there are fewer  scenarios for  a nonvanishing WEDM than for the AWMDM. For instance, neutral scalar bosons can only induce the tau WEDM  at the one-loop level as long as they are a mixture of $CP$ eigenstates. As for a  charged scalar boson, type-II contribution  vanishes for massless neutrinos and the same is true for type-I contribution, which is  nonvanishing only for two nondegenerate charged scalar bosons. We will not consider the case of doubly charged scalar bosons since, as we have seen above,  their couplings to charged leptons are too small to give a relevant contribution to the tau AWMDM, let alone to the WEDM. Another conclusion drawn from our study of the AWMDM is that multiple-Higgs doublet models are not expected to give a considerable enhancement to the weak properties of the tau lepton as compared to the contribution of a THDM.  Therefore, the most promising scenario for a nonvanishing WEDM of the tau lepton is that posed by a $CP$-violating THDM and we will refrain from analyzing other scenarios. We thus consider the scenario with two nondegenerate neutral scalar bosons that are a mixture of $CP$ eigenstates and evaluate the respective contributions as functions of their masses, which we  show in Fig. \ref{plotWEDMneutral}. Again these contributions could have a strong suppression once the appropriate coupling constants are introduced: for coupling constants of the order of $10^{-2}$, the values shown in the plots would decrease by about four orders of magnitude. In addition, stemming from the sum rules obeyed by the coupling constants, some of these contributions would be additionally suppressed  as the accompanying coupling constant could be small.

\begin{figure}[htb!]
\begin{center}
\includegraphics[width=17cm]{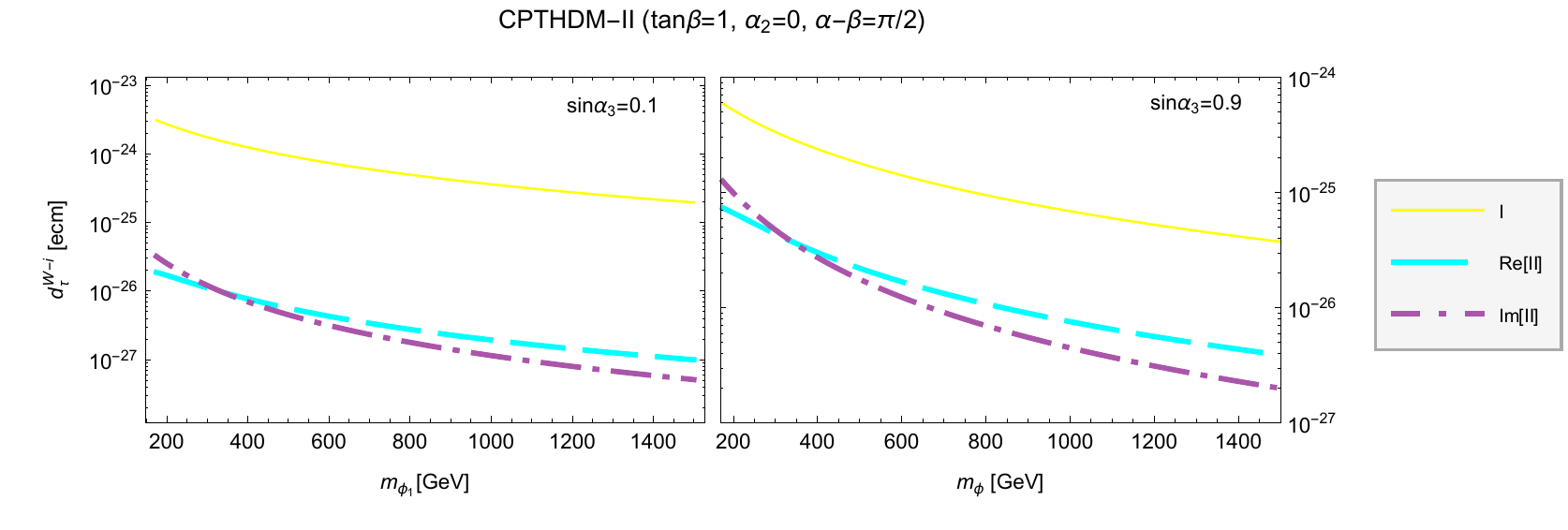}
\caption{Absolute values of the real  and imaginary parts of the partial  contributions from the $CP$-violating THDM-II to the WEDM of the tau lepton as functions of the scalar boson masses considering $m_{h_2}=200$ GeV and $m_{h_3}=m_\phi$ for type-I contribution as well as $m_{h_3}=m_{h_2}=m_\phi$ for type-II contribution. We use  the indicated values of the model parameters.  \label{plotWEDMCPTHDM}}
\end{center}
\end{figure}

Let us now consider the $CP$-violating THDM discussed above \cite{Gunion:1997aq} and assume the scenario with $\alpha_2=0$, in which $h_1$ is pure $CP$ even and it thus is identified with the SM Higgs boson, whereas $h_2$ and $h_3$ are mixtures of $CP$ eigenstates and they would give  the three types of contributions to the tau WEDM, though type-III contribution vanishes when $\alpha-\beta=\pi/2$.  We assume massless neutrinos so the contribution of  the charged scalar boson vanishes. The corresponding coupling constants are  shown Table \ref{CPVTHDMCouplings}, where in addition to $\alpha_2=0$, we set  $\alpha-\beta\simeq \pi/2$ and use $\alpha_1=\pi/2-\alpha$. We plot the partial contributions to $d_\tau^W$ from the two neutral scalar bosons $h_2$ and $h_3$ in Fig. \ref{plotWEDMCPTHDM} as functions of the scalar boson masses and for two values of $\alpha_3$. We can observe that  the dominant contribution to $d_\tau^W$ arises from the Feynman diagram of type-I, whereas diagrams of type-II give a negligible contribution, which however give the only nonvanishing contribution to the  imaginary part of $d_\tau^W$. In this scenario, the real part of the WEDM of the tau lepton is of the order of $10^{-24}$ ecm for small $\sin\alpha_3$ and decreases by almost one order of magnitude for a larger value, whereas the imaginary part is of the order of $10^{-26}$ for relatively light masses of the scalar bosons, but decreases quickly as $m_{h_3}$ increases.

\section{Conclusions and outlook}
\label{Conclusions}
We have performed an analysis of the contributions of models with an extended scalar sector to the tau AWMDM and WEDM, for which we have obtained analytic expressions both in terms of parametric integrals and Passarino-Veltman scalar functions. We first presented a model-independent analysis of the potential contributions of new neutral, singly charged, and doubly charged scalar bosons to the tau WDMs. Afterwards we focused on the contributions of some particular extension models. Although some of these contributions were studied before, prior to 2012, we have considered the most up-to-date constraints on the parameter space of each model, which take into account the  discovery of  a SM-like Higgs boson. It is found that the contributions of this kind of models, such as  multiple-Higgs doublet models or Higgs triplet models reduce  to those of THDMs, which arise from two new neutral and a singly charged scalar bosons, and are somewhat suppressed as compared to other extension models, which stems from the fact that the new scalar bosons typically have suppressed couplings to the SM particles. In particular, doubly charged scalar bosons have very suppressed coupling to the tau lepton in triplet models. Also, although HTMs predict the $H^\pm W^\mp Z$ vertex at the tree level, contrary to THDMs, its contribution to the tau WDMs is not relevant.

As far as the tau AWMDM is concerned, its real part reaches values as high as $10^{-10}-10^{-9}$ for masses of the new scalar boson in the 200 GeV range, but it decreases quickly as these masses increase. On the other hand, the imaginary part of $a_\tau^W$ is one or two orders of magnitude below. We also find that the most promising scenario for a nonvanishing WEDM of the tau lepton is that posed by $CP$-violating THDMs, which predicts three neutral scalar bosons $h_i$ $(i=1,2,3)$ that are a mixture of $CP$ eigenstates. This model allows for the presence of a nonvanishing $Zh_i h_j$ $(i\ne j)$ vertex, which is a novel prediction. We analyze a scenario where the lightest neutral scalar bosons is  $CP$-even and coincides with the SM Higgs boson. The dominant contribution is given by the neutral scalar bosons $h_2$ and $h_3$. The real part of $d_\tau^W$ is of the order of $10^{-24}$ ecm and its imaginary part can reach the $10^{-26}$ level for masses of the new scalar bosons of the order of a few hundred of GeVs.

In summary, the contributions of models with an extended scalar sector to the tau AWMDM and WEDM are smaller than those predicted by other types of extensions models. Although interesting by their own,  models with an extended scalar sector could  be the low-energy approximation of a more fundamental theory still unknown with a strongly interacting ultraviolet completion, which could give an enhancement to the WMDMs of a charged lepton. There are several models with  extended gauge sector that require  a scalar sector with additional scalar multiplets. Among the most popular ones are the MSSM, the left-right $SU(2)_R\times SU(2)_L\times U(1)_Y$ symmetric model and its supersymmetric version,  $SU(3)_L\times U(1)_X$ models, and little Higgs models, to mention a few ones. These models can include several new contributions to the tau WDMs arising from the new gauge bosons and fermions predicted by these theories. The results presented here can be useful to asses the magnitude of the contribution of the scalar sector of these models but we expect that the scalar boson contributions to the tau AWMDM are subdominant. On the other hand, the contribution to the WEDM can be relevant if there is no additional $CP$ violation sources in the respective extension model.

\acknowledgments{We acknowledge support from Conacyt and SNI (M\'exico). Partial support from VIEP-BUAP is also acknowledge. The work of G. Hern\'andez-Tom\'e was supported by Conacyt Project 236394 (Ciencia B\'asica)}.

\appendix
\section{Feynman rules}
\label{FeynRules}

\begin{figure}[htb!]
\begin{center}
\includegraphics[width=6cm]{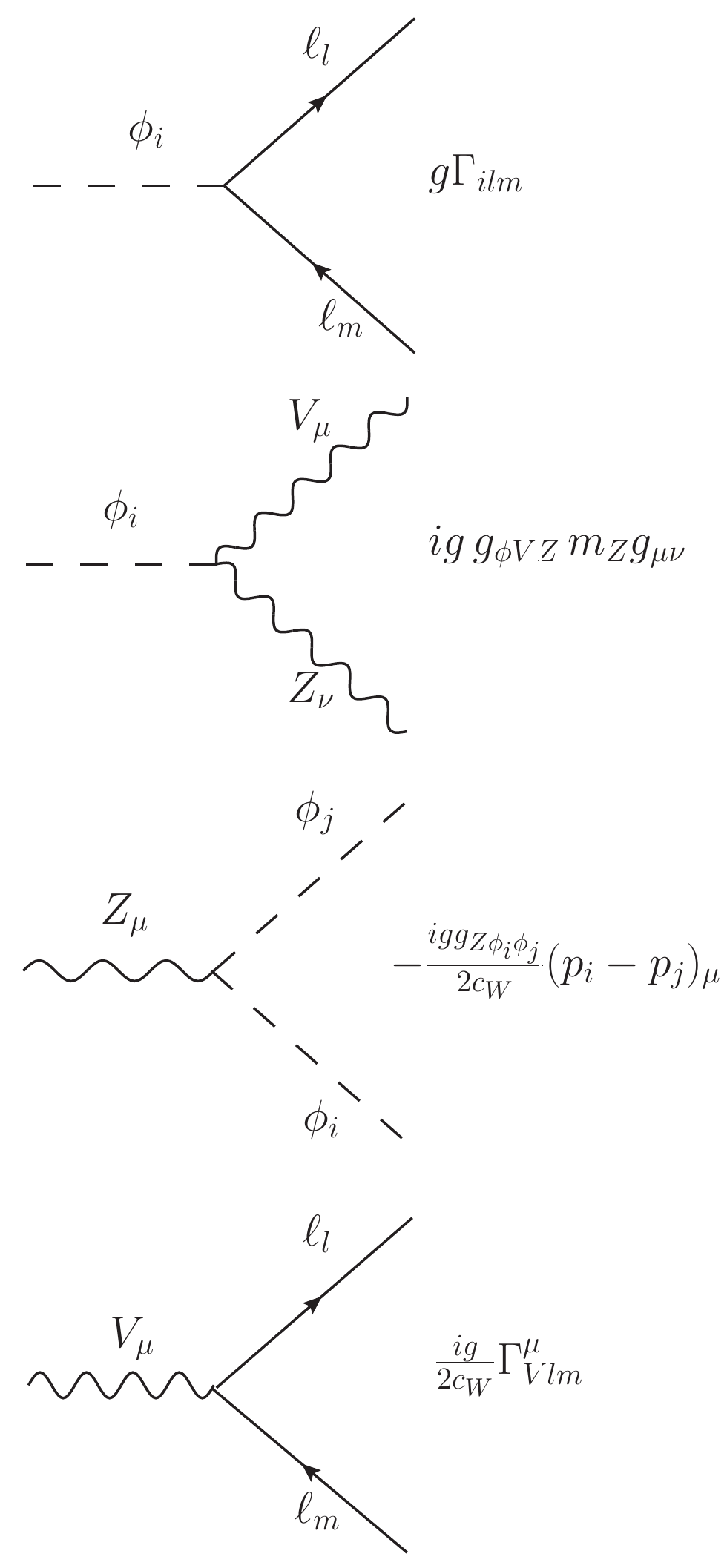}
\caption{Generic Feynman rules for the LNC interactions  necessary to calculate the weak properties of a charged lepton.  Here $\ell_l$ stands for a charged lepton and $\ell_m$ is a either a charged  or neutral  lepton, whereas the charges of the scalar bosons  $\phi_{i,j}$  and the gauge boson $V$ are fixed by charge conservation in each vertex. \label{FeynRules1}}
\par\end{center}
\end{figure}

In this appendix we present the Feynman rules necessary for the calculation of the static weak properties of a charged lepton.  In Fig. \ref{FeynRules1} we present the generic Feynman rules necessary when  LNC vertices are involved. The fermion, scalar boson, and  gauge boson propagators are the usual ones
and we refrain from presenting them here.

\begin{figure}[htb!]
\begin{center}
\includegraphics[width=6cm]{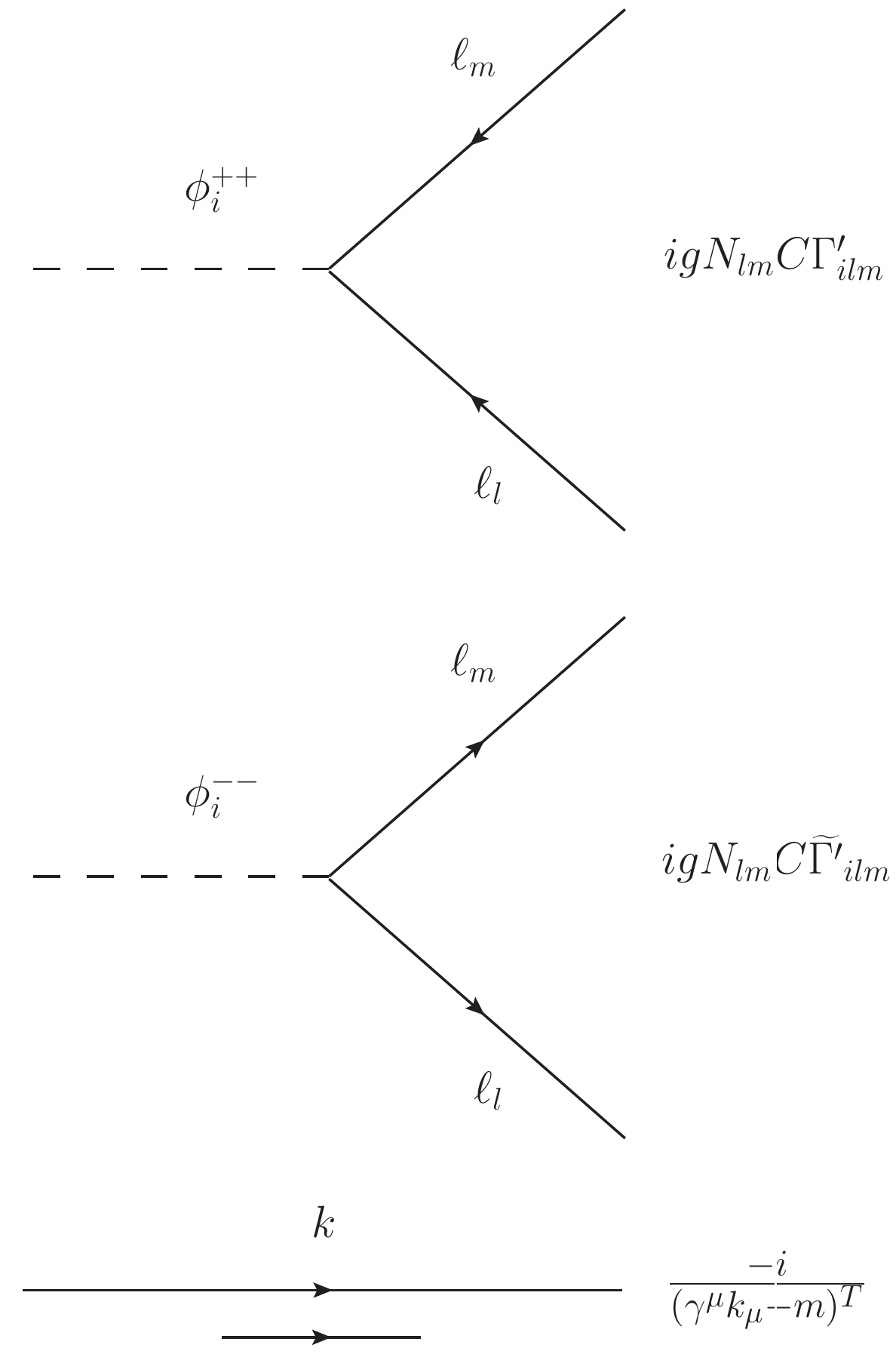}
\caption{Generic Feynman rules for the LNV interactions mediated by a doubly charged scalar necessary to calculate the weak properties of a charged lepton. Here $\ell_l$ and $\ell_m$ are both charged leptons. $C$ is the charge conjugation matrix, $\Gamma'_{ilm}=S'_{ilm}+P'_{ilm}\gamma^5$, and $\widetilde{\Gamma'}_{ilm}=C\gamma^0\Gamma^{'\dagger}_{ilm}\gamma^0C^{-1}= C\left(S^{'*}_{ilm}-P^{'*}_{ilm}\gamma^5\right)C^{-1}$.    $N_{lm}$ is a symmetry factor that is $2$ for $m=l$ and 1 otherwise. Note that the arrow below the fermion propagator stands for the direction in which the Feynman line is read, which in this case coincides with the fermion-flow direction.
\label{FeynRules2}}
\par\end{center}
\end{figure}

\begin{figure}[htb!]
\begin{center}
\includegraphics[width=6cm]{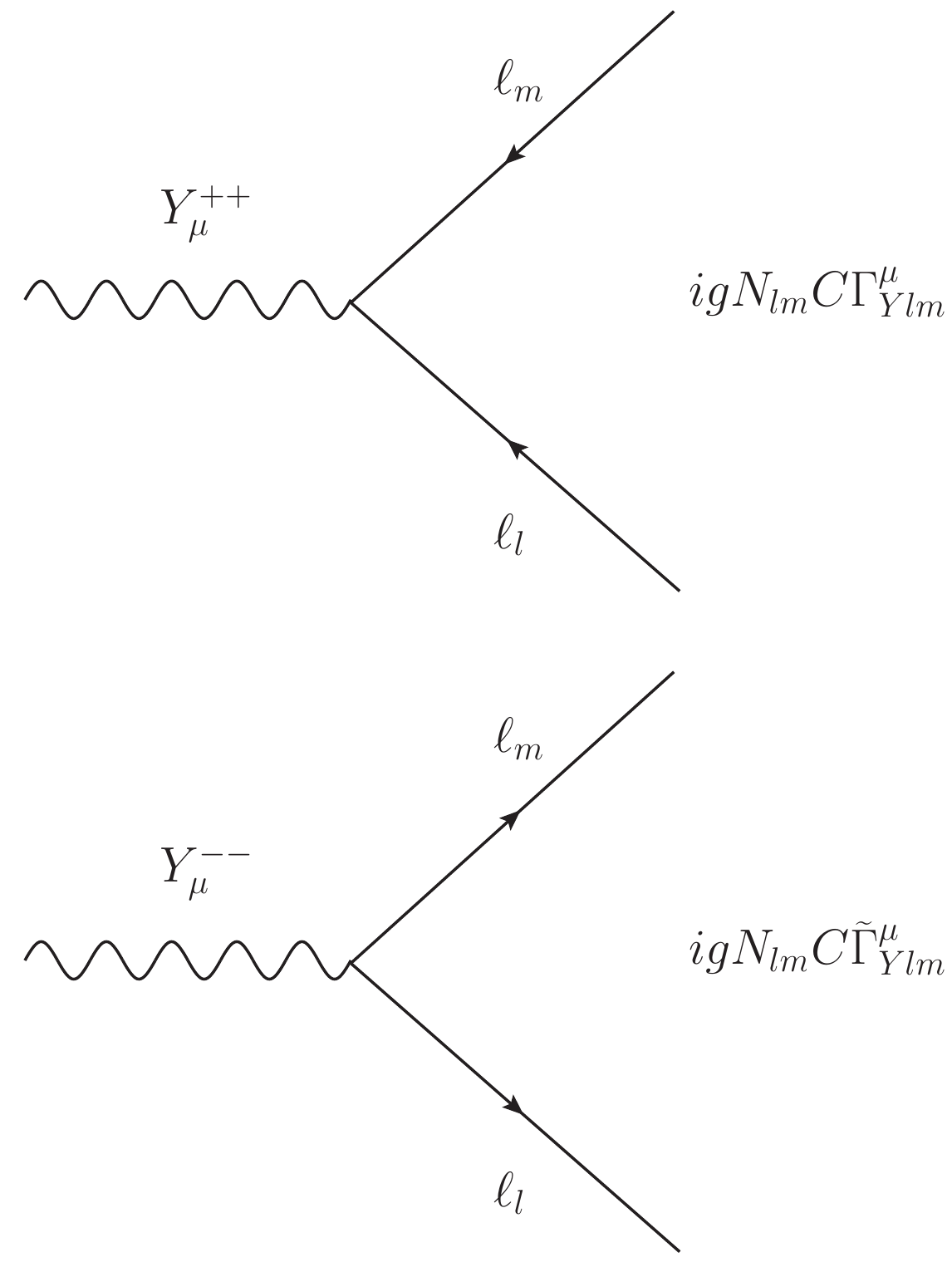}
\caption{The same as in Fig. \ref{FeynRules2}, but for the LNV interactions of a doubly charged gauge boson $Y$. Here   $\Gamma_{Ylm}^\mu=\left(g^{Y l m}_V-g^{Y l m}_A \gamma^5\right)\gamma^\mu$, $\widetilde{\Gamma}^\mu_{Ylm}=C\gamma^0\Gamma^{\mu\dagger}_{lm}\gamma^0C^{-1}=C\left(g^{Y l m}_V-g^{Y l m}_A \gamma^5\right)\gamma^\mu C^{-1}$.
\label{FeynRules3}}
\par\end{center}
\end{figure}

When  LNV vertices mediated by doubly charged scalar and gauge bosons are involved, we need the  Feynman rules  shown in Fig. \ref{FeynRules2}  and \ref{FeynRules3} . In calculating the amplitudes for these contributions we have followed the approach of  Ref. \cite{Moore:1984eg}) for the evaluation of Feynman diagrams with LNV vertices. To simplify the final amplitude we need to exploit the properties of the charge conjugation matrix such as $C=-C^{-1}$ and $C\gamma_\mu C^{-1}=\gamma_\mu^T$.

\section{One-loop functions}
\label{OneLoopFunctions}
In this appendix we present our results  for the $A^{ABC}_i$ and $D^{ABC}_i$  functions  involved in the calculation of the AWMDM and WEDM of a charged lepton presented in Sec. \ref{Calculation}. We will present analytic expressions in terms of both parametric integrals and Passarino-Veltman scalar integrals.

\subsection{Parametric integrals}
\label{ParamInteg}

The $A^{ABC}_i$ and $D^{ABC}_i$  functions can be cast in the form of a one-dimensional parametric integral as follows

\begin{equation}
A^{ABC}_{i}=\int_0^1 a^{ABC}_i(t)dt,
\label{Integrals}
\end{equation}
where the letters  in the superscript $ABC$  denote the dependence on the masses of the different  particles circulating into the loop (in fact $A$ is the particle that couples to both external lepton lines whereas both $B$ and $C$ couple to the $Z$ boson) and the subscript is used to denote distinct functions.   Although there is also dependence on the external lepton mass, we will omit such a dependence  in order to avoid cumbersome expressions, thus we use the short-hand notation  $a^{ABC}_i(t)\equiv a^{ABC}_i(t,x_{ l},x_A,x_B,\cdots)$.  Similar expressions hold  for  $D^{ABC}_{i}$ and $d^{ABC}_i(t)$.

\subsubsection{Weak magnetic dipole moment}
For the  type-I Feynman diagram  we have

\begin{equation}
a^{ m\phi_{i}\phi_{j}}_{I}(t)=\ t \left(
(t-1) \sqrt{x_{ l}}-\sqrt{ x_{ m}}\right)F^{ m\phi_{i}\phi_{j}}(t),
\end{equation}
with the following auxiliary functions
\begin{equation}
F^{ABC}(t)=f^{ABC}(t)+f^{ACB}(t),
\end{equation}
and
\begin{equation}
f^{ABC}(t)=\frac{1}{\xi^{ABC}(t)}\arctan\left[\frac{t-1 +x_B-x_C}{\xi^{ABC}(t)}\right],
\end{equation}
and
\begin{eqnarray}
\xi^{ABC}(t)&=&\Big[4 t \left((t -1)x_{ l}+x_A\right)-2(t-1)\left(x_B+x_C\right)-
(x_B-x_C)^2-(1-t)^2\Big]^{\frac{1}{2}}.
\end{eqnarray}

As long as the type-II Feynman diagram  is concerned, the corresponding  $a_i^{ABC}$ functions are
\begin{equation}
a^{\phi_{i} {m} {m}}_{II_1}(t)=\left(\sqrt{x_{ m}}+t
   \sqrt{x_{ l}}\right)(1-t)F^{\phi_i m  m}(t),
\end{equation}
and
\begin{equation}
a^{\phi_{i} {m} {m}}_{II_2}(t)=t(t-1) F^{\phi_i m  m}(t),
\end{equation}
whereas the function associated with the type-III Feynman diagrams, whose amplitudes  have been  added up, is
\begin{equation}
 a^{ {m}\phi_{i}V}_{III}(t)=f_{0}^{ {m}V}(t)+2 f_{1}^{ {m}\phi_{i}V}(t)F^{ {m}\phi_{i}V}(t)
 +f^{\phi_{i}V}_{2}(t)G^{ {m}\phi_{i}V}(t),
\end{equation}
with the auxiliary functions given as
\begin{equation}
f^{AB}_{0}(t)=2 (t-1) \left((1-3 t) g^{AB}(t)-2(1-2 t)\right),
\end{equation}
\begin{equation}
g^{AB}(t)=\log \left[t \left((t-1)
   x_{ l}+x_A\right)-(t-1)x_B\right],
\end{equation}
\begin{equation}
G^{ABC}(t)=g^{AB}(t)-g^{AC}(t),
\end{equation}
\begin{eqnarray}
f^{ABC}_{1}(t) &=& t^2 \left(2\left( \sqrt{x_{ l}}\sqrt{ x_A}+8 x_{ l}-5 x_A\right)+5\left(
   x_B+x_C\right)-7\right)\nonumber \\&+& t \left(4(x_A- x_{ l})-\left(4
   x_C+7\right)x_B +2 x_B^2+\left(2 x_C-5\right)
   x_C+5\right)\nonumber \\&+&3 t^3 \left(1-4 x_{ l}\right)-\left(1-x_B\right){}^2+x_C^2,
\end{eqnarray}
and
\begin{equation}
f^{AB}_{2}(t)=t \left(2(x_A- x_B+2)-3 t\right)-x_A-x_B-1.
\end{equation}

\subsubsection{Weak electric dipole moment}
As far as the contributions to the WEDM of a charged lepton are concerned, they are given through the following functions
\begin{equation}
d^{ m\phi_{i}\phi_{j}}_{I}(t)=2t
   \left(\left(x_{\phi _i} - x_{\phi _j}\right)\sqrt{x_{ l}} -\sqrt{
   x_{ m}}\right) F^{ m\phi_{i}\phi_{j}}(t)+  t\sqrt{x_{ l}}G^{ m\phi_{i}\phi_{j}} (t),
\end{equation}
\begin{equation}
d^{\phi_{i} {m} {m}}_{II}(t)=\sqrt{x_{ m}}(t-1) F^{\phi_i m  m}(t),
\end{equation}
and
\begin{equation}
 d^{ {m}\phi_{i}V}_{III}(t)=-f^{ {m}V}_{0}(t)+2h^{ {m}\phi_{i}V}_{1}(t)F^{ {m}\phi_{i}V}(t)
+h^{ {m}\phi_{i}V}_{2}(t)G^{ {m}\phi_{i}V}(t),
\end{equation}
where  the $F^{ABC}$, $G^{ABC}$ and $f^{AB}$ were defined above. The remaining auxiliary functions are
\begin{eqnarray}
h^{ABC}_{1}(t)&=&t^2 \left(7-18 x_{ l}+5(2 x_A-x_B-
   x_C))\right)\nonumber \\ &+&t \left(  2 x_C \left(\sqrt{x_{ l}}
  \sqrt{ x_A}+x_{ l}-x_C\right)+x_B \left(2 (2 x_C-\sqrt{x_{ l}}
  \sqrt{ x_A}- x_{ l})+7\right)  \right.\nonumber \\&+& \left.2 (\sqrt{x_{ l}}\sqrt{ x_A}+3
   x_{ l}-2 x_A-x_B^2)+5 (x_C-1)\right)\nonumber \\&-&3t^3
   \left(1-4 x_{ l}\right)+\left(1-x_B\right){}^2-x_C^2,
\end{eqnarray}
and
\begin{equation}
h^{ABC}_{2}(t)=1+x_B+x_C+3 t^2-2 t \left(x_{ l}+\sqrt{x_{ l}}\sqrt{ x_A}+x_B-x_C+2\right).
\end{equation}

\subsection{Passarino-Veltman scalar functions}
\label{Passarino-Veltman}
We now present our results  in terms of Passarino-Veltman scalar functions. We first introduce the following set of ultraviolet finite scalar integrals

\begin{eqnarray}
\Delta_1 &=&B_0(0,m_A^2,m_A^2)-B_0(0,m_B^2,m_B^2),\\
\Delta_2 &=&B_0(0,m_B^2,m_B^2)-B_0(0,m_C^2,m_C^2),\\
\Delta_3 &=&B_0(0,m_C^2,m_C^2)-B_0(m_l^2,m_A^2,m_B^2),\\
\Delta_4 &=&B_0(m_l^2,m_A^2,m_B^2)-B_0(m_l^2,m_A^2,m_C^2),\\
\Delta_5 &=&B_0(m_l^2,m_A^2,m_C^2)-B_0(m_Z^2,m_A^2,m_B^2),\\
\Delta_6 &=&B_0(m_l^2,m_A^2,m_C^2)-B_0(m_Z^2,m_B^2,m_C^2),\\
\Delta_7 &=&m_Z^2C_0(m_l^2, m_l^2, m_Z^2, m_B^2, m_A^2, m_C^2).
\end{eqnarray}

\subsubsection{Weak magnetic dipole moment}

The $A^{ABC}_i$ functions are given by
\begin{eqnarray}
A^{ABC}_{I}&=&\frac{1}{4\rho _l}\Bigg(x_B-2 x_A+x_C+2 x_l -2 x_A\Delta_1
+\left(x_B-2 x_A\right)\Delta_2+\left(x_B-2 x_A+x_C\right)\Delta_3
\nonumber\\&-&\frac{1}{\delta _l}\Big( x_l \left(2\left(x_A+(2 x_B +1)x_l+4 \sqrt{x_A x_l}-2 x_B\right)+1\right)-2 \sqrt{x_A x_l}+x_A+x_C \left(2x_l(1-2 x_l)-1\right)\Big)\Delta_4
\nonumber\\&-&\frac{2 }{\delta _l}\left(x_l \left(3(2x_A- x_B- x_C)+2 x_l+1\right)+ 2\sqrt{x_A x_l}(4x_l-1)\right)\Delta_6
\nonumber\\&-&\frac{2}{\delta _l}\Big(2 x_l^2 \left(2 x_A-(x_B-x_C)^2+x_B+x_C-1\right)+x_l \left(6 x_A \left(x_B+x_C\right)-2x_A(3 x_A+2)\right.\nonumber\\&-&\left.x_B(4x_C+x_B-1)-x_C(x_C-1)\right)+\left(1+(x_B+x_C)(4x_l-1)+2x_l(4(x_l-x_A)-3)+2x_A\right)
\nonumber\\&\times&\sqrt{x_A x_l}+2 x_l^3\Big)\Delta_7
\Bigg),
 \end{eqnarray}
with $\delta_l=1-4x_l$ and $\rho _l=\delta_l x_l$,
\begin{eqnarray}
A^{ABC}_{II_1}&=&\frac{1}{4\rho _l}\Bigg(x_B-x_A-x_l+x_B\Delta_1+\left(x_B-x_A\right)\Delta_3
-\frac{1}{\delta _l}\left(2(3x_Ax_l- \delta_l\sqrt{x_A x_l})-x_l \left(6 x_B+2 x_l+1\right)\right)\Delta_6
\nonumber\\&-&\frac{1}{\delta _l}\Big(2  \delta _l \sqrt{x_Ax_l} \left(x_B-2x_A+x_l\right)-2x_l\left( x_A \left(6 x_B+2 x_l+1-3x_A\right)+\left(x_l(1-x_l-2 x_B)+x_B \left(3 x_B+2\right)\right)\right)\Big)\nonumber\\&\times&\Delta_7
\Bigg),\nonumber\\
 \end{eqnarray}

\begin{eqnarray}
A^{ABC}_{II_2}&=&\frac{1}{2\rho _l}\Bigg(x_A-x_B+x_l -x_B\Delta_1
+\left(x_A-x_B\right)\Delta_3-\frac{x_l}{\delta _l}\left(6 (x_B-x_A)+2 x_l+1\right)\Delta_6
\nonumber\\&-&\frac{2x_l}{\delta _l}\left((2 x_B-x_A+x_l) \left(3 x_A+x_l-1\right)-3 x_B^2\right)\Delta_7
\Bigg),
 \end{eqnarray}
and
\begin{eqnarray}
A^{ABC}_{III}&=&\frac{1}{\rho _l}\Bigg( \left(\sqrt{x_l}-\sqrt{x_A}\right)\Big(\left(2 x_A-x_B-2 x_l-x_V\right)+2 x_A \Delta_1+\left(2 x_A-x_B\right)\Delta_2
+ \left(2 x_A-x_B-x_V\right)\Delta_3\Big)
\nonumber\\&+&\frac{1}{\delta _l}\Big(\left(\sqrt{x_l}-\sqrt{x_A}\right)\left(x_A-2 \delta _l \sqrt{x_A x_l}+x_l(2 x_A + \left(4 x_B \left(x_l-1\right)+2 x_l+1\right))\right)
\nonumber\\&-&x_V \left(\sqrt{x_A} \left(2x_l(1-2 x_l)-1\right)+\sqrt{x_l} \left(2x_l(3-14 x_l)+1\right)\right) \Big)\Delta_4
\nonumber\\&+&\frac{2\sqrt{x_l}}{\delta _l}\left(x_l \left(2 (x_A+x_l)+3 x_B-5 x_V-1\right)+3 \sqrt{x_A x_l} \left(x_B-2 x_A+x_V+2x_l-1\right)+2 (x_A- x_V)\right)\Delta_6
\nonumber\\&+&\frac{2 \sqrt{x_l}}{\delta _l}\Big(x_V \left(2x_B-3 x_A+x_l(6 \left(2 x_B+1\right) x_l-8 x_B+3(2x_A-1))\right)\nonumber\\&+&x_A \left(2 x_l\left(x_B+1-2x_l\right) +x_B-1+2x_A \left(x_l-1\right)\right)\nonumber\\&+& \sqrt{x_Ax_l} \big(\left(x_B-1\right)^2+ 2x_V(x_V \left( x_l+1\right)+  \left(1-2 x_B\right) \left(x_l-1\right)+3x_A)+6x_A(x_A+\left(1-x_B-2 x_l\right))\nonumber\\&+&2 x_l\left(x_B(x_B+1)-2+3x_l\right)\big)+x_l \left(x_B(1-x_B) \left(2 x_l+1\right)+2 x_l \left(x_l-1\right)+x_V^2 \left(14 x_l-5\right)\right)\Big)\Delta_7
\Bigg).
 \end{eqnarray}

\subsubsection{Weak magnetic dipole moment}
As for the $D^{ABC}_i$ functions, they  are given by
\begin{eqnarray}
D^{ABC}_{I}&=&\frac{2}{\rho _l}\Bigg(\frac{1}{2}\delta _l^2 x_B \left(x_B-x_C\right)\Delta_2+\delta _l \left(x_B-x_C\right)\Delta_3
-\Big(x_l \left(4 x_A+2 x_B+\delta _l-6x_C\right)+2 \sqrt{x_A x_l}+x_C-x_A\Big)\Delta_4
\nonumber\\&-&4 \left(\sqrt{x_A x_l}+x_l(x_B-x_C)\right)\Delta_5
+2\Big((\sqrt{x_A x_l}+x_l \left(x_B-x_C\right)) \left(1+2 x_A-x_B-x_C\right)\nonumber\\&-&2x_l (\sqrt{x_A x_l}-x_l \left(x_C-x_B\right))\Big)\Delta_7
\Bigg),
 \end{eqnarray}

\begin{eqnarray}
D^{ABC}_{II}&=&\frac{16 \sqrt{x_Ax_l}}{\rho _l}\left(- \Delta_6
+\left(x_B-x_A+x_l\right)\Delta_7
\right),
 \end{eqnarray}
and
\begin{eqnarray}
D^{ABC}_{III}&=&\frac{1}{\rho _l x_V}\Bigg(\delta _l \left(\sqrt{x_A}+\sqrt{x_l}\right)\left(x_V-x_B\right)\Big)-x_B \delta _l \left(\sqrt{x_A}+\sqrt{x_l}\right)\Delta_2
+\delta _l \left(\sqrt{x_A}+\sqrt{x_l}\right) \left(x_V-x_B\right)\Delta_3
\nonumber\\&+&\Big(x_V \left(\sqrt{x_A} \left(1-6 x_l\right)+\sqrt{x_l} \left(2 x_l+1\right)\right)+\left(\sqrt{x_A}+\sqrt{x_l}\right) \left(x_l \left(4 x_A+2 x_B+1-4x_l\right)-2 \sqrt{x_Ax_l}-x_A\right)\Big)\Delta_4
\nonumber\\&-&2 \sqrt{x_l} \left(\sqrt{x_Ax_l} \left(x_V-x_B+1\right)+x_A-x_V+x_l (x_V-x_B)\right)\Delta_6
\nonumber\\&+&2\sqrt{x_l} \Big(x_B \left(2 \left(x_V-\sqrt{x_A x_l}(1+x_A-x_l)\right)-x_l \left(2 (x_A+2x_V- x_l)+1\right)-x_A+x_B \left(\sqrt{x_A x_l}+x_l\right)\right)\nonumber\\&+&\sqrt{x_A x_l} \left(x_V+1\right) \left(2 (x_A-x_l)-x_V+1\right)+x_l \left(2 x_A \left(x_V-1\right)+x_V \left(3 x_V-1-2x_l\right)\right)\nonumber\\&+&x_A \left(2 x_A-3 x_V+1\right)\Big)\Delta_7
\Bigg).\nonumber\\
 \end{eqnarray}

\bibliography{biblio}

\end{document}